\documentclass[12pt,preprint]{aastex}
\usepackage{amssymb}
\usepackage{amsmath}
\usepackage{graphicx}

\newcommand\kms{km~s$^{-1}$}

\newcommand\co{$^{12}$CO}
\newcommand\mjb{mJy~beam$^{-1}$}
\newcommand\jb{Jy~beam$^{-1}$}

\newcommand\pp{$^{\prime\prime}$}
\newcommand\um{$\mu$m}

\newcommand\q{$\sim$}

\newcommand\h{H$_{2}$}
\newcommand\msun{M$_{\odot}$}  

\newcommand\lsun{L$_{\odot}$}

\newcommand{\ammonia}{NH$_3$}

\newcommand{\hco}{HCO$^{+}$}

\newcommand\water{H$_{2}$O}\defcitealias{maserpap}{C09}

\newcommand\ujb{$\mu$Jy~beam$^{-1}$}
\newcommand{\meth}{CH$_3$OH}

\newcommand{\vlsr}{$v_{LSR}$}
\newcommand{\degree}{$^{\circ}$}
\newcommand{\whz}{W Hz$^{-1}$}

\defcitealias{egocat}{C08} 	
\defcitealias{maserpap}{C09} 	
\defcitealias{C11}{C11}

\begin{document}
\shortauthors{Cyganowski et al.}

\title{Deep Very Large Array Radio Continuum Surveys of GLIMPSE Extended Green Objects (EGOs)}
\author{C.J. Cyganowski\altaffilmark{1,4}, C.L. Brogan\altaffilmark{2},
  T.R. Hunter\altaffilmark{2}, E. Churchwell\altaffilmark{3}}

\email{ccyganowski@cfa.harvard.edu}

\altaffiltext{1}{Harvard-Smithsonian Center for Astrophysics, Cambridge, MA 02138}
\altaffiltext{2}{NRAO, 520 Edgemont Rd, Charlottesville, VA 22903}
\altaffiltext{3}{University of Wisconsin, Madison, WI 53706}
\altaffiltext{4}{NSF Astronomy and Astrophysics Postdoctoral Fellow}

\begin{abstract}

We present the results of deep, high angular resolution Very Large
Array (VLA) surveys for radio continuum emission towards a sample of 14
GLIMPSE Extended Green Objects (EGOs).  Identified as massive young
stellar object (MYSO) outflow candidates based on their extended 4.5
\um\/ emission in \emph{Spitzer} images, the EGOs in our survey sample
are also associated with 6.7 GHz Class II and/or 44 GHz Class I
\meth\/ masers.  No continuum is detected at 3.6 or 1.3 cm towards the
majority (57\%) of our targets (median rms \q 0.03 and 0.25 \mjb ).
Only two EGOs are associated with optically thin emission consistent
with ultracompact/compact HII regions.  Both of these sources exhibit
cm$-\lambda$ multiplicity, with evidence that one of the
less-evolved members may be driving the 4.5 \um\/ outflow.  Most of
the other cm-$\lambda$ EGO counterparts are weak ($\lesssim$ 1 mJy),
unresolved, undetected at 1.3 cm, and characterized by intermediate
spectral indices consistent with hypercompact (HC) HII regions or
ionized winds or jets.  One EGO cm counterpart, likely an optically
thick HC HII region, is detected only at 1.3 cm and is associated with
hot core line emission and \water\/ and 6.7 GHz \meth\/ masers.  The
results of our exceptionally sensitive survey indicate that EGOs
signify an early stage of massive star formation, before photoionizing
feedback from the central MYSO significantly influences the
(proto)cluster environment.  Actively driving outflows (and so,
presumably, actively accreting), the surveyed EGOs are associated with
significant clump-scale gas reservoirs, providing sufficient material
for sustained, rapid accretion.

\end{abstract}


\section{Introduction}\label{intro}
A key difference between high-mass stars and their low-mass
counterparts is that massive stars are sufficiently luminous to emit
ionizing radiation.  Early efforts to compile large samples of massive
young stellar objects (MYSOs)\footnote{We define MYSOs as young
stellar objects (YSOs) that will become main sequence O or early-B
type stars (M$>$8 \msun).}  focused on ultracompact (UC) HII regions,
identified by their cm-$\lambda$ radio continuum emission and
observationally defined as having sizes $\le$ 0.1 pc, densities $\ge$
10$^{4}$ cm$^{-3}$, and emission measures $\ge$ 10$^{7}$ pc cm$^{-6}$
\citep[e.g.][]{WC89,KCW94,Hoare07}.  The main mechanisms proposed to explain why UC HII regions are not always observed
around luminous ($\gtrsim$10$^{4}$ \lsun) MYSOs--''quenching'',
gravitational trapping, and swelling--are consequences of the high
accretion rates associated with massive star formation
\citep[e.g][]{Walmsley95,Keto07,HoareFranco07,Yorke08,Hosokawa09,Hosokawa10}.
This led \citet{HoareFranco07} to suggest that UC
HII regions appear when accretion ceases, though
observations of ionized accretion flows indicate that the
transition is not clear-cut
\citep[e.g.][]{Keto02,KetoWood06,KetoKlaassen08,Roberto09}.  While an
evolutionary sequence for MYSOs is still under development, it is
generally accepted that a UC HII region represents the photoionization
of its environment by a young O or early B type star, and thus a late
stage of massive star formation
\citep[e.g.][]{ZY07,Ellingsen07,Breen10}.  A good example of this
stage is G5.89$-$0.39, where the candidate ionizing star has been
identified \citep{Feldt03} in the midst of a dust-free cavity
\citep{Hunter08}.

The first appearance of observable emission from ionized gas
associated with a MYSO is considerably less well understood.
Hypercompact (HC) HII regions, ionized accretion flows, outflows,
jets, stellar winds, and ionized/photoevaporating disks all may exist
around forming massive stars, but the relationships among these
phenomena, including when each ``turns on,'' remain the subject of
considerable discussion
\citep[e.g.][]{MvdT04,Keto07,GibbHoare07,HoareFranco07,KetoKlaassen08,Rodriguez08,Roberto09}.
HC HIIs are defined by their scale: size $\lesssim$ 0.05 pc, density
$\gtrsim$ 10$^{6}$ cm$^{-3}$ \citep[e.g.][]{Hoare07,Lizano08}.  An
additional distinction is that sources in which photoionization by the
central MYSO dominates, including ionized accretion flows or disks,
are considered HC HIIs.  In contrast, in jets and winds, shocks may
contribute significantly to the ionization \citep[see for
example][]{Shepherd04,Keto07,Roberto10}.  HC HIIs, winds, and jets can
be difficult to distinguish observationally, as all are characterized
by intermediate cm-$\lambda$ spectral indices (S$_{\nu} \propto
\nu^{\alpha}$, $-$0.1$<\alpha<$ $+$2) and all are generally fainter
than UC HIIs at cm wavelengths
\citep{Sewilo04,Kurtz05,GibbHoare07,Keto08,Lizano08}.  Thus the
distinctions among these phenomena are in large part dynamical: a key
observational discriminant is the recombination line width (or for
jets, proper motion), with jets having the largest
velocities/linewidths, followed by winds, then HC HII and finally UC
HII regions \citep{Hoare07,HoareFranco07}.
 
Despite the uncertainties outlined above, the strong association
between ongoing accretion (and hence youth) and absence of a UC HII
region means that characterizing the presence/absence of cm-$\lambda$
continuum emission is a key step in understanding the evolutionary
state of any sample of MYSOs.  A large new sample of MYSOs \emph{with
active outflows}--and so presumably ongoing accretion--has been
identified from \emph{Spitzer Space Telescope} surveys of the Galactic
Plane based on extended 4.5 \um\/ emission.  Known as ``Extended
Green Objects (EGOs)'' \citep{egocat,maserpap} or ``green
fuzzies''\citep{Chambers09} for the common coding of three-color IRAC
images, their extended 4.5 \um\/ emission is thought to arise from
molecular line emission shock-excited in protostellar outflows
\citep[predominantly
\h:][]{SmithRosen05,Smith06,Davis07,Ybarra09,Ybarra10,DeBuizer10}.
\citet[hereafter C08]{egocat} cataloged over 300 EGOs in the Galactic Legacy
Infrafred Mid-Plane Survey Extraordinaire (GLIMPSE-I) survey
\citep{Churchwell09}, and argued that their MIR properties were consistent with young, embedded MYSOs.  To test the
hypothesis that GLIMPSE EGOs are specifically \emph{massive} YSOs with
\emph{active} outflows, \citet[hereafter C09]{maserpap} conducted a VLA search for
two diagnostic types of \meth\/ masers towards EGOs: 6.7 GHz Class II
\meth\/ masers, associated exclusively with MYSOs
\citep[e.g.][]{Minier03,Bourke05,Xu08,Pandian08}, and 44 GHz Class I
\meth\/ masers, associated with molecular outflows and outflow-cloud
interactions \citep[e.g.][]{PlambeckMenten90, Kurtz04}.  The detection
rates were remarkably high, confirming the MYSO and outflow nature of
the EGO targets: of a sample of 28 EGOs, $\gtrsim$ 64\% have
associated 6.7 GHz \meth\/ masers, and \q 89\% of the 6.7 GHz maser
sources also have associated 44 GHz masers.  A
complementary JCMT survey found evidence for warm
dense gas associated with EGOs \citepalias{maserpap} and \citet[hereafter C11]{C11} found hot cores and
high-velocity, bipolar molecular outflows in two EGOs studied in
detail with high-resolution mm 
observations.  All of this is strong evidence for EGOs being young,
actively accreting MYSOs, but constraints on the cm-$\lambda$
continuum properties of EGOs are conspicuously lacking.  The 44 GHz
continuum observations of \citetalias{maserpap} ruled out bright UC HII
regions as powering sources for most (95\% of) targeted EGOs, but the
sensitivity was insufficient to detect fainter UC HII or HC HII
regions.  In this paper, we present deep, high-resolution VLA 3.6 and
1.3 cm continuum observations of 14 EGOs from the \citetalias{maserpap}
sample, with the aim of constraining the evolutionary state of EGOs.

\section{Very Large Array Observations and Data Reduction}\label{vlaobs}

We observed a sample of 14 EGOs with the Very Large Array (VLA)\footnote{The
National Radio Astronomy Observatory operates the VLA and is a
facility of the National Science Foundation operated under agreement
by the Associated Universities, Inc.} at 3.6 cm (8.46 GHz) and 1.3 cm
(22.46 GHz) using the standard continuum mode (2 $\times$ 50 MHz, dual
polarization, for an effective bandwidth of 172 MHz\footnote{http://www.vla.nrao.edu/astro/guides/exposure/}).  The 3.6
cm data were taken in the VLA B configuration on 2009 May 7 and 14.
Each day, short observations of each target were staggered throughout
the observing block to optimize the uv-coverage, for a total on-source
time of \q45 minutes per source.  Calibration followed standard
procedures in AIPS, including the use of a model for 3C286, the
absolute flux calibrator.  The quasars J1751+096, J1733-130,
J1925+211, and J1911-201 were used as phase calibrators and to correct
for closure errors on EVLA-VLA baselines using the AIPS task BLCAL.
We estimate that the absolute flux calibration is accurate to \q5\%,
and that the absolute positional uncertainty is \q0\farcs1.

We chose the VLA C configuration for our 1.3 cm observations in order
to approximate as closely as possible the uv coverage and synthesized
beamsize (\q1\pp) of the B configuration 3.6 cm data.  The 1.3 cm data
were taken on 2009 July 10 and 17 using fast switching (cycle time 2.5
minutes) and reference pointing.  The 22 GHz zenith opacity was \q
0.16-0.17 on July 10 and \q 0.13-0.14 on July 17.  Cycles on each
source were distributed across the observations to improve
uv-coverage; the total on-source integration time was \q25-30 minutes
per target.  Table~\ref{obstable} lists the fast-switch phase
calibrators.  Observations of the bright quasars J1751+096 and
J2253+161 were used to correct for EVLA-VLA closure errors using the
AIPS task BLCAL.  The data were calibrated in AIPS following standard
high-frequency procedures, including using a model for 3C286.  For the
July 17 data, 3C286 was used for absolute flux calibration.  On July
10, the 3C286 data were unusable, so the absolute flux scale was set
using J2253+161, assuming S(22.46 GHz)=6.22 Jy based on VLA
monitoring.\footnote{http://www.aoc.nrao.edu/\raise.17ex\hbox{$\scriptstyle\sim$}smyers/calibration/2009/K\_band\_2009.shtml}
Comparison of the derived fluxes for the phase calibrators and
J1751+096 between our datasets and with VLA flux monitoring shows no
systematic offset, and suggests the absolute flux calibration is
accurate to \q10\%.

The data were imaged in AIPS, and, for sources with sufficient
signal-to-noise, self-calibrated.  
For each target, we also
convolved the 3.6 cm image to the 1.3 cm resolution, to facilitate
direct comparison of fluxes and limits between the two datasets.
Image parameters for each EGO are presented in Table~\ref{obstable}.
Some fields are affected by image artifacts from poorly-sampled
large-scale emission.  In these cases, images were made first with all data
to search for extended emission (\S\ref{discussion_evol}), and then with uv
limits to reduce artifacts and improve sensitivity to compact
emission; uv weighting adjustments are indicated in
Table~\ref{obstable}.  The largest angular scale of the observations
is \q 20\pp, but we cannot recover fluxes for objects more extended
than \q 10\pp.  The primary beam (FWHM) of the VLA is 5.3\arcmin\/ at
3.6 cm and 2.0\arcmin\/ at 1.3 cm.  All measurements were made from
images corrected for the primary beam response.

\section{Results}\label{results}

Of the 14 EGOs in our sample, 6 are detected in cm continuum emission.
We define a cm-$\lambda$ EGO ``detection'' as $>$4$\sigma$ 
emission within the polygonal aperture for the EGO published
by \citetalias{egocat}.  Table~\ref{egos_table_detect} lists observed properties
of cm continuum sources; for EGOs associated with multiple cm sources,
they are designated -CM1, -CM2, etc., in order of descending peak
intensity.  The vast majority of detected cm sources are unresolved at
the \q1\pp\/ resolution of our observations.  For sources detected
with sufficient signal-to-noise, a single two-dimensional Gaussian was
fit, and the fitted centroid position, peak intensity, integrated flux density,
and deconvolved source size are reported in Table~\ref{egos_table_detect}.
For weaker sources for which the size is not well-constrained,
Table~\ref{egos_table_detect} lists the fitted centroid position and peak intensity
from a Gaussian fit assuming an unresolved point source.  
For marginal detections ($<$ 5$\sigma$), the position and intensity
of the peak pixel are reported in Table~\ref{egos_table_detect}, and the
quoted uncertainty is the 1$\sigma$ rms in the image.  The parameters of the
one resolved, irregular cm source (G49.27$-$0.34-CM1) were measured
using the polygon aperture photometry program developed for the
CORNISH survey \citep{CORNISH}\footnote{http://www.ast.leeds.ac.uk/$\scriptstyle\sim$phycrp/software.html}, and
Table~\ref{egos_table_detect} reports the position of the brightness-weighted
source center, the peak intensity, the integrated flux density, and the equivalent source diameter.
Table~\ref{egos_table_nondetect} lists 4$\sigma$ upper limits for EGOs not
detected in cm continuum emission.  Figure~\ref{3color_zoom} shows the
spatial relationship of the MIR, cm-$\lambda$ continuum, and Class I
and Class II \meth\/ maser emission for EGOs with cm-$\lambda$
detections.  For all sources except G11.92$-$0.61 (which is detected
only at 1.3 cm), contours of the 3.6 cm image are shown in
Figure~\ref{3color_zoom}.  Most EGO cm counterparts are detected only
at 3.6 cm; for sources that are detected at both
wavelengths, the 1.3 cm images (not shown) have similar morphology,
but lower signal-to-noise.

In addition to the EGO counterparts listed in Table~\ref{egos_table_detect},
we detect other continuum sources within the VLA field of view that
are not ``associated'' with the target EGO (as defined above).  Such
sources are listed in Table~\ref{others_table} if they are within
a 1\arcmin\/ radius of the VLA pointing center (corresponding to the
FWHM of the VLA primary beam at 1.3 cm), so that a 1.3 cm flux or
limit can be measured.  Figure~\ref{3color_big} shows three-color IRAC
images of a 2\arcmin\/ square field, centered
on the VLA pointing center, for each of the EGO fields in
Table~\ref{others_table}, overlaid with contours of the VLA 3.6 cm,
MIPSGAL 24 \um\/ \citep{Carey09}, and Bolocam Galactic Plane Survey
\citep[BGPS;][]{Aguirre11,Rosolowsky10} 1.1 mm continuum emission.  The
information in Table~\ref{others_table} is equivalent to that in
Table~\ref{egos_table_detect}, and, for compact sources, measured as
described above.  For extended sources poorly imaged by the
interferometer, only estimated positions, sizes, and angular
separations from the target EGO are listed in
Table~\ref{others_table}.
The sources in Table~\ref{others_table} are designated ``F'' (for
``field''), the name of the targeted EGO, and -CM1, -CM2, etc. in
order of descending peak intensity.  
The angular separations of these
sources from the target EGO positions range from 8-51\pp.  Some are
likely associated with the same star-forming region as the EGO, and
are noted in Section~\ref{individ_sources}.  In general, however,
insufficient information is available to determine association and/or
the physical properties of these sources, and we do not include them
in our analysis.

\subsection{Physical Properties: Estimates and Limits from the Cm-$\lambda$ Continuum}\label{phy_prop_results}

For EGOs undetected at both 3.6 and 1.3 cm, we adopt two approaches to constraining the physical properties of the (outflow) driving source and of any ionized gas.  The number of ionizing photons corresponding to an unresolved 4$\sigma$ source is given \citep[following][]{KCW94} by:
\begin{equation}\label{nc_thin}
N^{\prime}_{C} < 7.59 \times 10^{48} \, \,  T_e^{-0.5} {\left(\frac{1}{\alpha(\nu,T_e)}\right)}{\left(\frac{\nu}{GHz}\right)}^{0.1} {\left(\frac{S_{\nu}}{Jy}\right)} {\left(\frac{D}{kpc}\right)}^2
\end{equation} 
where $N^{\prime}_{C}$ has units of (photons) s$^{-1}$, T$_{e}$ is the
electron temperature in K, $\alpha(\nu,T_e)$ is a factor of order
unity, interpolated from the tables in \citet{Mezger67}, and units for
the other quantities are as indicated.  This calculation
provides an upper limit to the ionizing radiation that could be
emitted by the MYSO and yet be undetected in our observations.  However, the upper limit obtained from
equation~\ref{nc_thin} may be an underestimate, as it assumes optically thin, ionization-bounded free-free emission and no
absorption of ionizing radiation by dust \citep[see for example
discussion in][]{Keto08}.
  
If, in contrast, the emission is assumed to be optically thick, an
upper limit to the source size can be derived from \begin{math}
S_{\nu}=B_{\nu}(T_e)\Omega_s \end{math} where S$_{\nu}$ is the
observed limit on the flux density, B($\nu,T_{e}$) is the Planck
function, and $\Omega_s$ is the source solid angle.  In the
Raleigh-Jeans limit,
\begin{equation}\label{thick_radius_limit}
r[AU] < 6.639 \times 10^5 {\left[\frac{S_{\nu} D^2}{T_e \nu^2} \right]}^{0.5}
\end{equation}
where r is the radius (assuming spherical geometry), S$_{\nu}$ is the
flux density (limit) in Jy, D is the distance in kpc, T$_{e}$ is the
electron temperature in K, and $\nu$ is the observing frequency in
GHz.  Recent studies assume electron temperatures in the range
8000-10000 K for HC and UC HII regions
\citep[e.g.][]{Pandian10,Battersby10,Katharine09}.  The difference in
the values of the derived parameters is $<$15\%.  We adopt T$_{e}$=10000 K here.

For all EGOs undetected at both 3.6 and 1.3 cm in our observations, Table~\ref{nondetect_table} lists the limits on the source size and
ionizing photon rate.  The 3.6 cm data are significantly more sensitive, with a
median 4$\sigma$ limit on N$_{Lyc}$ of 2.24$\times$10$^{44}$
s$^{-1}$, compared to 1.81$\times$10$^{45}$ s$^{-1}$ for the 1.3 cm
data.  We include both estimates because the turnover frequency is a function
of density \citep[e.g.][]{Kurtz05}, so emission from dense HII regions
is more likely to be optically thin at shorter wavelengths.  
However, emission from HC HII regions may be optically thick even
at 1.3 cm, and \citet{Keto08} discuss ways in which density gradients within HII
regions may ``stretch'' the transition region of the
spectrum.

For EGOs detected at either 3.6 or 1.3 cm, Table~\ref{detect_table}
lists the 3.6 to 1.3 cm spectral index or limit, and the estimates for
the ionizing photon flux and source size calculated from equations
\ref{nc_thin} and \ref{thick_radius_limit} and the flux densities in
Table~\ref{egos_table_detect}.  Spectral index limits are calculated
using the 4$\sigma$ upper limit at the undetected wavelength.  The size
estimates from equation \ref{thick_radius_limit}, which assume optically thick emission, are omitted in cases
where the spectral index indicates the emission is optically thin
($\alpha \sim -$0.1), but retained for sources with intermediate
spectral indices for comparison.  For all sources, physical properties
were calculated assuming the distances from \citetalias{maserpap},
which are listed in Tables~\ref{nondetect_table} and
\ref{detect_table}.  Throughout \S\ref{individ_sources}, spectral types corresponding to N$_{Lyc}$ are
taken from \citet{Smith02} for stars earlier than B1.5V, and
from \citet{Doyon90} for later B-type stars.

\subsection{Notes on Individual Sources}\label{individ_sources}

\subsubsection{G10.29$-$0.13}\label{g1029}
This EGO is unique among the \citetalias{maserpap} sample in being
associated with both 6.7 GHz Class II and 44 GHz Class I \meth\/
masers, but lacking a discrete MIPS 24 \um\/ counterpart.  The detectability
of a weak 24 \um\/ counterpart is, however, limited by
the point-spread function (PSF) wings of the adjacent
MIR-bright complex.  This
saturated MIPS 24 \um\/ source, located east of the EGO, is
detected at both 3.6 cm and 1.3 cm in our VLA images
(Table~\ref{others_table}, Fig.~\ref{3color_big}).  However, the cm-$\lambda$
emission is extended, and not well-imaged.
Lower-resolution (37$\times$25\pp) 20 cm VLA images show that this cm source is
part of an extended ionized complex, designated G10.30$-$0.15 \citep{KimKoo01}.  The velocity
of the molecular gas associated with this ionized complex \citep[v$_{LSR}$=13.5 \kms,][]{KimKoo03} agrees
well with the radio recombination line velocity reported for
G10.315$-$0.150 \citep[12 \kms,][]{Downes80}, and with the \vlsr\/
of the dense gas associated with the EGO \citepalias[13.6 \kms,][]{maserpap}, indicating that the EGO
is part of the same star-forming region.  

\subsubsection{G11.92$-$0.61}\label{g1192}

The EGO G11.92$-$0.61 is the only source in our VLA continuum survey
to be detected only at 1.3 cm.  The 1.3 cm detection of 1.07 \mjb\/
is marginal (\q 4.2 $\sigma$), but intriguing because it implies a
spectral index $>$1.7.
As shown in Figure~\ref{3color_zoom}, the 1.3 cm continuum emission is
coincident with the 1.4 mm dust continuum emission from the hot core
MM1.  The 1.3 cm peak is \q 0\farcs4
($\gtrsim$ 1500 AU) southwest of the CARMA position of MM1
\citepalias{C11} and a \water\/ maser \citep{HC96}, which are
coincident within astrometric uncertainties.  The cm peak
is \q 0\farcs2 ($\gtrsim$ 930 AU) northwest of the intensity-weighted
6.7 GHz Class II \meth\/ maser position from
\citetalias{maserpap} (Fig.~\ref{3color_zoom}).  
The 1.3 cm detection is too strong to be pure dust emission, as this
would account for at most 0.2 mJy (extrapolating from the comparable
resolution CARMA data with a spectral index of 3).
If the 1.3 cm emission is due to
optically thick free-free emission ($\alpha$=2) and remains optically
thick into the mm, then free-free emission could potentially account
for \q59\% of the CARMA 1.4 mm flux of MM1, with the remainder due to
dust emission.  The calculated ionizing photon flux, N$_{Lyc}\ge$1.66$\times$10$^{45}$ s$^{-1}$, would
correspond to a single ionizing ZAMS star of spectral type B2.  However, there is strong evidence that
G11.92$-$0.61-MM1 is actively driving an outflow, including Class I
\meth\/ masers and SiO, 4.5 \um, and high-velocity \co(2-1) and
\hco(1-0) emission \citepalias{C11,maserpap}--and therefore actively accreting.  It is
therefore very unlikely that the driving source is in fact in a ZAMS
configuration
\citep[e.g.][]{HoareFranco07,Yorke08,Hosokawa09,Hosokawa10}.

\subsubsection{G18.67+0.03}\label{g1867}

The EGO G18.67+0.03 is the easternmost of four MIPS 24 \um\/ sources
associated with a ridge of 1.1 mm emission (Fig.~\ref{3color_big}).
The three eastern 24 \um\/ sources are all associated with 44 GHz
Class I \meth\/ masers; two of these sources, including the EGO, are
also associated with 6.7 GHz Class II \meth\/ masers.  (The
westernmost 24 \um\/ source falls just outside the field searched for
maser emission by \citetalias{maserpap}.)
Interestingly, the only cm continuum emission in the field is detected
towards the 24 \um\/ source associated with Class I but not Class II
\meth\/ masers.  Maser evolutionary sequences have generally placed
Class I \meth\/ masers among the earliest observable signs of massive
star formation, due largely to their association with molecular
outflows and outflow/cloud interactions
\citep[e.g.][]{Ellingsen07,Breen10}.  Indeed, one of the brightest
known Class I maser sources is NGC 6334 I(N), which contains a rich
cluster of compact millimeter continuum sources \citep{Brogan09}.
However, \citet{Voronkov10} have recently suggested that Class I
\meth\/ masers may also be excited by expanding HII regions driving
shocks into the surrounding molecular cloud, and so (also) trace a
much later stage of massive star formation.  Because most
high-resolution Class I maser searches to date have targeted Class II
masers, there are few Class I \meth\/ maser sources associated with
cm-$\lambda$ continuum emission \emph{and known to lack Class II
\meth\/ masers} \citep{Voronkov10}.  The cm continuum source F
G18.67+0.03-CM1 is an important addition to this sample.  Since the
focus of this paper is on the cm-$\lambda$ properties of EGOs, we
defer further analysis of F G18.67+0.03-CM1 to a future
paper that will present high resolution mm-$\lambda$ molecular line
and continuum data, which are required to better characterize the
physical properties and relative evolutionary states of the EGO and
the other 24 \um\/ sources. 

\subsubsection{G19.36$-$0.03}\label{g1936}

Two weak ($\lesssim$ 1 mJy) 3.6 cm sources are detected in this field,
neither coincident with the EGO.  The
stronger cm source, F G19.36$-$0.03-CM1, is \q 8\farcs0 ($\gtrsim$
19400 AU) northwest of the nominal EGO position cataloged by
\citetalias{egocat}, and \q4\farcs5 ($\gtrsim$10900 AU) northwest of
the intensity-weighted 6.7 GHz \meth\/ maser position \citepalias{maserpap}.  This cm source is
coincident with multiband IRAC and with MIPS 24 \um\/ emission; a short line of 44
GHz Class I \meth\/ masers originates near the cm source and extends \q 2\pp\/ NNW.
F G19.36$-$0.03-CM1 is thus another candidate for Class I \meth\/
masers possibly excited by an expanding HII region.  The NW-SE
extension of the MIPS 24 \um\/ emission indicates that it is likely a
blend of emission from multiple sources, possibly F G19.36$-$0.03-CM1
and the MYSO pinpointed by the 6.7 GHz \meth\/ maser. 
These sources may be two members of a (proto)cluster, possibly in different evolutionary
stages.  We defer further discussion of F
G19.36$-$0.03-CM1 to a future paper that will present high-resolution
mm-$\lambda$ line and continuum data to better constrain the
properties of the (proto)cluster members.  The nature of the weaker cm
source, F G19.36$-$0.03-CM2, and its relation (or lack thereof) to the
EGO are unclear.  F G19.36$-$0.03-CM2 is coincident with compact,
multiband
IRAC emission, but lacks strong 24 \um\/ emission
(Fig.~\ref{3color_big}), and could be a much more evolved object.

\subsubsection{G24.94+0.07}\label{g2494}
The 3.6 cm EGO counterpart EGO G24.94+0.07-CM1 is coincident with both
6.7 GHz Class II \meth\/ maser and MIPS 24 \um\/ emission
(Fig.~\ref{3color_zoom}).  The fitted position of the cm source
is offset by only \q0\farcs4 ($\gtrsim$ 1100
AU) from the intensity-weighted 6.7 GHz maser position and by
\q1\farcs4 ($\gtrsim$ 3400 AU) from the nominal MIPS 24 \um\/ position
\citepalias{maserpap}.  The nondetection
at 1.3 cm implies $\alpha$ $<$0.7.
The 3.6 cm emission is unresolved, and 
the ionizing radiation flux 
is low (4.51$\times$10$^{44}$ s$^{-1}$, ZAMS star of
type B2 to B3).  Since the central MYSO is actively driving an
outflow--as indicated by the 4.5 \um\/ emission, Class I \meth\/
masers, and SiO emission--and accreting, however, it is unlikely to
have contracted to a ZAMS configuration (see also \S\ref{g1192}).  

\subsubsection{G25.27$-$0.43}\label{g2527}

Two weak ($\lesssim$ 1 mJy) 3.6 cm sources are detected in this field,
both $>$0.5 pc in projection from the EGO.  The weaker, F
G25.27$-$0.43-CM2, is coincident with the center of a nebula that is
bright in all IRAC bands and at 24 \um\/ (Fig.~\ref{3color_big}), and corresponds to the
\emph{IRAS} source 18362-0703 (\q 28\pp\/ south of the EGO).  The
brighter cm source, F G25.27$-$0.43-CM1, does not appear to have a MIR
counterpart, and its nature is unclear.

\subsubsection{G28.28$-$0.36}\label{g2828}

A weak 3.6 cm source, EGO G28.28$-$0.36-CM1, is marginally detected
(\q 4.1$\sigma$) \q 1\farcs2 ($\gtrsim$ 3800 AU) northeast of the 6.7
GHz \meth\/ maser associated with the EGO \citepalias[intensity-weighted
maser position,][]{maserpap}.  
The limit on the spectral index ($<$1.6) is unconstraining.

Three other cm sources are detected in the field.  F EGO
G28.28$-$0.36-CM1 is the well-studied UC
HII region G28.28$-$0.36
\citep[e.g.][]{Walsh98,Longmore07,Churchwell10}.  
\citet{Churchwell10} report a
H30$\alpha$ recombination line velocity of 40.8 \kms\/
(FWHM 31.5 \kms) and a \hco(3-2) velocity of 47.1 \kms\/ towards the UC HII region.  The latter is in good agreement with the
molecular gas velocity measured towards the EGO 
\citepalias[49.5 \kms,][]{maserpap}.  The EGO and the UC HII region ($\gtrsim$
0.5 pc away) are thus likely
part of the same star-forming complex, which may also include the compact cm source F
G28.28$-$0.36-CM3 (coincident with a MIR nebula \q 24\pp $\gtrsim$ 0.4
pc southeast of the UC HII region).  The other field source, F EGO
G28.28$-$0.36-CM2, does not have a MIR counterpart.

\subsubsection{G28.83$-$0.25}\label{g2883}

This EGO lies at the edge of the MIR bubble N49 \citep{bubblescat}, and several authors have investigated the
possibility of triggered star formation around the bubble rim
\citep[e.g.][]{Watson08,Zavagno10,Deharveng10}.  The 3.6 cm emission
\q13\pp\/ ($\gtrsim$ 0.3 pc) north of the EGO (F G28.83$-$0.25-CM1,
over-resolved in our VLA image)
corresponds to a compact HII region seen in MAGPIS 20 cm data.  \citet{Deharveng10} quote a 20 cm flux density $\le$ 0.1 Jy and a spectral
type for the ionizing star of BOV.  The CO(3-2) maps of \citet{Beaumont10} show a velocity
gradient of only a few \kms\/ in the molecular gas around the bubble
rim, confirming that the three 24 \um\/ sources in
Figure~\ref{3color_big} \citepalias[the compact HII region, the EGO, and
the 24 \um\/ source south of the EGO, see also][]{maserpap} are part of
the same star-forming clump.

Two faint ($<$ 0.5 mJy) 3.6 cm continuum sources are detected
coincident with the EGO.  The nature of the brighter source, EGO
G28.83$-$0.25-CM1, is somewhat puzzling: located \q5\farcs2
($\gtrsim$26200 AU) west of the 6.7 GHz Class II \meth\/ maser
position \citepalias{maserpap}, CM1 is similarly offset from the 24 \um\/ peak
and appears to coincide with a deficit of 4.5 \um\/ emission
(Fig.~\ref{3color_zoom}).  G28.83$-$0.25-CM1 has an intermediate
spectral index ($<$1.1) and modest ionizing photon flux
(7.95$\times$10$^{44}$ s$^{-1}$).
The fainter 3.6 cm continuum source, EGO
G28.83$-$0.25-CM2 (\q 4.7$\sigma$), is coincident with the 6.7 GHz
\meth\/ maser and with 24 \um\/ emission.  The spectral index is poorly constrained 
($<$1.9), and the size implied in the case of optically thick emission
is small (\q 90 AU).

\subsubsection{G35.03+0.35}\label{g3503}

The EGO G35.03+0.35 is remarkable among our sample for its
cm-$\lambda$ multiplicity: five compact 3.6 cm sources are detected, with projected
separations $\lesssim$20000 AU (Fig.~\ref{3color_zoom}).  In addition, \citet{rsro} find
evidence for an additional source NW of CM1 and CM2, detected in
\ammonia\/ and 1.3 mm continuum emission, but not at cm wavelengths.
CM1, CM2, and CM3 all lie along a ridge of 24 \um\/
emission.  The MIPS image is saturated, so the position
of the 24 \um\/ peak is highly uncertain \citepalias[see also][]{maserpap}; however,
the peak of CM2 lies within the saturated MIPS pixels,
while the peaks of the other cm sources do not.  CM2 is also
coincident with blueshifted 6.7 GHz Class II \meth\/ masers
\citepalias [$\Delta$v \q 6-12 \kms,][]{maserpap}, blueshifted OH masers
\citep[$\Delta$v \q 3-14 \kms,][]{Argon00}, redshifted \water\/ masers
\citep[$\Delta$v \q 13-17 \kms,][]{FC99}, and hot molecular gas seen in
\ammonia(6,6) emission \citep[T$_{k}$=220 K,][]{rsro}.

CM1 was detected in the \citet{KCW94} survey of UC HII regions, and also at 7 mm by
\citetalias{maserpap}.  
To better constrain the SED of CM1, we
reimaged the 7 mm continuum data presented by \citetalias{maserpap}, applying a uvtaper
to approximate as closely as possible the beam of our 3.6 and 1.3 cm
data.  We then convolved the tapered 7 mm image with the beam listed in
Table~\ref{obstable}, and corrected for the more recent VLA flux scale
used for the 3.6 and 1.3 cm data.  The flux density of CM1
measured from this new 7 mm image 
is 10.9$\pm$3.0 mJy (including a
20\% uncertainty in the absolute flux calibration).  In
addition, we measured the 6 cm flux density of CM1 from the publicly
available CORNISH image\footnote{Available at
http://www.ast.leeds.ac.uk/Cornish/public/release1.php}
\citep[][]{CORNISH}, and find a 6 cm flux density of 10.7$\pm$0.4 mJy. 
We fit a free-free emission model as described in \citet{Hunter08} to
the 6 cm-7 mm data (Fig.~\ref{g3503_seds}), obtaining best-fit parameters
N$_{e}$=9.5$\times$10$^{4}$ cm$^{-3}$, T$_{e}$=10400 K, and size
0\farcs4 (0.007 pc\q1370 AU).  This size agrees well with the deconvolved source sizes from two-dimensional Gaussian fits to our 3.6 and
1.3 cm images (\q 1230$\times$1100 AU and \q 1750$\times$1060 AU, respectively).
\citet{rsro} find an electron temperature of 7900 K and a density of
1.3$\times$10$^{4}$ cm$^{-3}$ from analyzing H63$\alpha$ and
H64$\alpha$ recombination line emission towards CM1; the lower values likely represent a contribution from a
more extended, less dense component in the lower resolution EVLA data
(beam 3\farcs7 $\times$ 3\farcs0, fitted source size 1\farcs3).  While
its size would place CM1 in the HC HII regime, the electron
density and modest recombination line width \citep[17.6
\kms,][]{rsro} are characteristic of UC HII regions
\citep[e.g.][]{Kurtz05,Hoare07,Murphy10}.  The ionizing photon flux derived from our 3.6 and 1.3 cm observations (\q 1.5 $\times$10$^{46}$ s$^{-1}$) is consistent with a single ionizing star of spectral type B1.5V.

While the deconvolved sizes of CM1 and CM2 are quite similar (CM2: \q
1750$\times$820 AU), their spectra are very
different.  As shown in Figure~\ref{g3503_seds}, CM2 has a rising
spectrum between 3.6 cm and 7 mm (it is undetected in the 6 cm CORNISH
image, 4$\sigma$ limit 1.48 mJy).  The spectral index between 3.6 and
1.3 cm is 0.67$\pm$0.14, derived from a Monte Carlo calculation of the
spectral index that accounts for the uncertainties in the 
flux densities.  The measured 7 mm flux density (7.5$\pm$2.2 mJy,
from the tapered and convolved image described above) is higher than
predicted by extrapolation with $\alpha_{3.6\_1.3cm}$
(Fig.~\ref{g3503_seds}).  This 7 mm excess may be due to dust emission
\citep[see also][]{rsro}.
Alternatively, \citet{Keto08} have suggested that HC HII SEDs
with similar features can be fit purely as ionized emission
from HII regions with steep density gradients.
Models of clumpy nebulae can also reproduce
intermediate spectral indices for HC HII regions
\citep[e.g.][]{IgnaceChurchwell04}.
In general, an intermediate
spectral index indicates that a range of opacities is present within
a source.  The deconvolved size of CM2 is significantly
larger than the sizes derived assuming optically thick emission
(Table~\ref{detect_table}).
(Sub)arcsecond-resolution (sub)mm data are needed to constrain the
relative contributions of dust and free-free emission in CM2.

The natures of the other cm sources are only loosely constrained by our
observations.  The coincidence of CM3 with the NE extension of the 24
\um\/ emission suggests that it may be a MYSO; however,
\citet{DeBuizer06} found that the 12 and 18 \um\/ MIR emission from
the MYSO (and EGO) G35.2$-$0.74 was dominated by emission from warm
dust in the outflow cavity walls.  The 24 \um\/ ridge has
approximately the same NE-SW axis as the bipolar 4.5 \um\/ lobes, so
the possibility that the unsaturated 24 \um\/ emission NE and SW of
CM2 traces an outflow cavity cannot be discounted.  In this case, the
\q symmetric spacing of CM3 and CM4 relative to CM2 might be suggestive of knots in an ionized jet.
In any case, at least three (and possibly more) MYSOs are clustered at
$\lesssim$ 20000 AU scales, suggestive of a proto-Trapezium such as
seen in an increasing number of mm (proto)clusters
\citep[e.g.][C11]{Hunter06,C07,Rodon08}.  \citet{Sewilo04} also found that
all the HC HII regions in their sample were in pairs or multiples with
other cm-$\lambda$ sources, though generally with wider separations
($>$ 0.2 pc).

\subsubsection{G49.27$-$0.34}\label{g4927}

Two cm continuum sources are detected coincident with the EGO, one
resolved source that is detected at both 3.6
and 1.3 cm, and an unresolved source 
detected only at 3.6 cm.  The morphology of the resolved source, EGO G49.27$-$0.34-CM1, is complex
(Fig.~\ref{3color_zoom}), with a bright, compact ``head'' (\q2\farcs2
$\times$ 1\farcs2 \q 0.06 $\times$ 0.03 pc), a knotty ring-like
structure (\q5\farcs1 $\times$ 4\farcs6 \q 0.14 $\times$ 0.12 pc), and
more extended diffuse emission 
(dimensions measured from the 3.6 cm
image).  The 3.6 and 1.3 cm flux densities 
are consistent with optically thin free-free
emission; 
the ionizing photon flux 
(\q 2.2$\times$10$^{47}$ s$^{-1}$) corresponds to a single ionizing star
of spectral type B0V, in agreement with the estimate
of \citet{Mehringer94} based on an unresolved 20 cm detection (\q
14\pp\/ resolution).
The spectral index limit for EGO G49.27$-$0.34-CM2 ($<$ 0.2) is also
consistent with optically thin free-free emission, but the calculated
ionizing photon flux is about two orders of magnitude lower
(1.79$\times$10$^{45}$ s$^{-1}$).  

Both cm
sources are coincident with MIPS 24 \um\/ emission; notably, neither
is coincident with the 44 GHz Class I \meth\/ masers detected by
\citetalias{maserpap}. 
This is of interest
because \citet{DeBuizer10} obtained Gemini L and M band spectra for
this EGO, and detected only continuum emission (no \h\/ or CO).  Their
slit crossed the bright multiband IRAC and MIPS 24 \um\/ source near
CM2, and passed near the northern 44 GHz maser \citep[Fig.2
of][]{DeBuizer10}.  The lack of \h\/ emission in the Gemini spectrum
and the 20 cm detection by \citet{Mehringer94} made this EGO a
candidate for Class I \meth\/ masers excited by shocks from expanding
HII regions \citep{Voronkov10}, but the
large spatial offset complicates this interpretation.  An outflow--perhaps driven by CM2, or by a MYSO undetected
at cm wavelengths--may in fact be present, and responsible for the
shocks exciting the maser emission.  In this picture, shocked \h\/ or CO in the outflow would contribute to the 4.5
\um\/ emission to the south, near the Class I
masers, but was either outside the \citet{DeBuizer10} slit or too
weak to be detected in the Gemini spectrum.

\subsubsection{G49.42+0.33}\label{g4942}
Two 3.6 cm sources are detected in the field, one (F G49.42+0.33-CM1)
coincident with a multiband IRAC source and 24 \um\/ peak and only \q
0\farcs3 ($\gtrsim$ 4100 AU) from the eastern 6.7 GHz \meth\/ maser (G49.417+0.324) detected by
\citetalias{maserpap}.  The other source, F G49.42+0.33-CM2, is extended
and poorly imaged in our high-resolution VLA data; it is coincident
with 24 \um\/ emission and an 8 \um\/ nebula.  If they are at the same
distance as the EGO, the cm sources are $\gtrsim$ 0.7 pc (F
G49.42+0.33-CM1) and $\gtrsim$ 1.8 pc (F G49.42+0.33-CM2) from the EGO
position.  It is unclear whether the cm
sources are part of the same molecular cloud/star-forming complex as
the extended 4.5 \um\/ source.  This EGO is also by far the most
distant in our sample, so the cm nondetection at the EGO position
corresponds to limits for the ionizing photon flux about an order of
magnitude higher than for the other targets
(Table~\ref{nondetect_table}).

\section{Discussion}\label{discussion}

\subsection{Detection Rate: Comparison with Other Studies}\label{rate_dis}

As stated in \S\ref{results}, we detect continuum emission at either
3.6 or 1.3 cm from 6/14 (\q 43\%) of the EGOs in our VLA survey sample.
Including the four EGOs
from the \citetalias{maserpap} sample with deep continuum observations in
the literature, all nondetections, brings the detection rate to 6/18 (\q 33\%).\footnote{G23.96$-$0.11,
G37.48$-$0.10, and G39.10+0.49 are undetected at 3.6 cm in VLA A-configuration
observations: 1$\sigma$ rms \q 50 \ujb\/ (\q 60\% higher
than our median), $\theta_{syn}$ \q 0\farcs35 $\times$
0\farcs25 \citep{Bart09}.  
G23.01$-$0.41 is undetected at 1.3 cm: 4$\sigma$ limit \q 0.5 \mjb (half our median), $\theta_{syn}$ \q
1\pp\/ \citep{Codella97}.
We downloaded and reduced archival 3.6 cm VLA data (B configuration, project code AH868) for the remaining \citetalias{maserpap} EGO, G10.34$-$0.14.  Unfortunately, imaging is compromised by the poor uv coverage of these data, and the rms is not constraining.} 
Since this sample is predominantly
a 6.7 GHz \meth -maser-selected EGO subsample, the most relevant comparisons are searches for cm-$\lambda$ continuum
emission towards Class II \meth\/ masers.  Two such recent
studies have sensitivities and detection limits roughly comparable to
our deep VLA survey.  \citet{Pandian10} surveyed 20 6.7
GHz \meth\/ masers selected from a blind single-dish maser survey, and
\citet{Bart09} targeted 30 6.7 GHz \meth\/ masers selected
from both blind and \emph{IRAS}-targeted single dish surveys.  The
\citet{Pandian10} and \citet{Bart09} samples include 2 and 4 EGOs from
the \citetalias{egocat} catalog, respectively, and 3 of the EGOs observed by \citet{Bart09} are also in the \citetalias{maserpap} sample.

\citet{Pandian10} obtained shallow 3.6 cm and much deeper 1.3 cm VLA observations
(both with $\theta_{syn}$ \q 1\pp); at 1.3 cm, they detect 30\% (6/20) of the targeted masers in
cm continuum emission.  
The \citet{Pandian10} data and our VLA 1.3 cm observations provide
comparable limits on central source properties; while their 1.3 cm
observations are roughly twice as deep (1$\sigma$ rms 0.12 \mjb),
their targets are significantly more distant (median D=10 kpc,
compared to 3.8 kpc for our EGO sample).
\citet{Bart09} detect 3.6 cm continuum emission towards \q 13\% (4/30,
with one additional questionable association) of their 6.7 GHz \meth\/
maser sample, 
the distance range of which \citep{Bart11} is similar to that of our target EGOs. 
Notably,
in three sources \citet{Bart09} detect weak ($\lesssim$ 1 mJy), unresolved or marginally resolved 3.6 cm
emission with the continuum peak offset $<$ 0\farcs2 from the
6.7 GHz maser, similar to the continuum detections towards
several EGOs (Fig.~\ref{3color_zoom}, \S\ref{individ_sources}).

Considering our 1.3 and 3.6 cm detections separately for straightforward
comparison, we detect \q31\% of EGOs at 3.6 cm (4/13, 4/16 \q 25\% including
literature data) and \q15\% of EGOs at 1.3 cm (2/13, 2/14 \q 14\% including
literature data). 
(These numbers exclude G49.27$-$0.34, the only EGO in our sample without associated 6.7 GHz \meth\/
masers.)
In sum, the detection rate for 1.3 cm continuum emission towards EGOs
from the \citetalias{maserpap} sample is lower than that of
\citet{Pandian10} towards 6.7 GHz \meth\/ masers, though the limiting
sensitivities in physical terms (e.g. ionizing photons s$^{-1}$) are
similar.  In contrast, the detection rate for 3.6 cm continuum
emission towards these EGOs is higher than that of
\citet{Bart09} for their 6.7 GHz \meth\/ maser sample.  Considering
the small number statistics for the detections, however, the detection
rates for all three studies are roughly comparable.

\subsection{Nature of Cm Continuum Emission}\label{lir_dis}

A number of
diagnostic plots for distinguishing ionized emission from UC HII
regions, HC HII regions, and MYSO winds and jets have been proposed \citep{Hoare07,HoareFranco07}.
One key discriminant is the recombination
line width, but such measurements are
not available for most of our target EGOs.  We can place our
sources in L$_{\nu}$(8 GHz) versus L$_{bol}$(24 \um)\footnote{The
quantity plotted is the total bolometric luminosity, referred to by
\citet{Hoare07} and \citet{HoareFranco07} as L(IR).  We estimate
bolometric luminosities from 24 \um\/ flux densities, and so use the
term L$_{bol}$(24 \um).} parameter space, albeit roughly, by using
MIPS 24 \um\/ fluxes as a proxy for bolometric luminosity.  24 \um\/ is the wavelength closest to the
SED peak for which we currently have images with sufficient angular
resolution to avoid source confusion.

\citet{Mottram11} have explored using the \emph{MSX} 21 \um\/
broadband flux as a proxy for bolometric luminosity for YSOs and HII
regions in the Red \emph{MSX} Source (RMS) sample.  For 613
sources with bolometric luminosities determined by fitting well-sampled
SEDs, they find a mean F$_{bol}$/F$_{MSX 21 \mu m}$=21.9$\pm$1.9.  While
the ratio does not depend on
source type (YSO v. HII region), the scatter is more than an order
of magnitude (\q 7-200).  For the MIPS 24 \um\/ filter, the same
analysis of the RMS sample gives F$_{bol}$/F$_{MIPS 24
\mu m}$=29.5$\pm$1.9 (J. Mottram, priv. comm.), with comparable scatter (\q 8-150).  There is, however, some evidence that
this mean ratio may underestimate the bolometric luminosities of EGOs.

G19.01$-$0.03 is a rare example of an EGO in which the ``central''
source is a GLIMPSE point source, clearly resolved from extended
emission in the IRAC images.  \citetalias{C11} observed this EGO at
\q10,000 AU resolution in 1.3 mm continuum emission with the SMA, and
detect only a single compact core, suggesting the observed MIR-mm
emission arises from a single, embedded MYSO with a bolometric
luminosity of \q 10$^{4}$ \lsun\/ derived from SED modeling
\citep{Robitaille06,Robitaille07}.
Extrapolating from the 24 \um\/ flux density using the
mean F$_{bol}$/F$_{MIPS 24 \mu m}$ ratio from the RMS data gives a
bolometric luminosity of only 1.6$\times$10$^{3}$ \lsun .  It is reasonable to suppose that
the SEDs of EGOs might peak at longer wavelengths than the MIR-bright
population of YSOs and HII regions included in the RMS sample.  If EGOs represent a younger stage, then the volume of surrounding dust that the central (proto)star has been able to heat will be much smaller than for an HII region, and so the average temperature will be lower.
In this case, one would expect the RMS F$_{bol}$/F$_{MIPS 24 \mu m}$ ratio
to underpredict the bolometric
luminosities of EGOs.  For G19.01$-$0.03, a range in
F$_{bol}$/F$_{MIPS 24 \mu m}$ of 8 to 150 corresponds to calculated
bolometric luminosities of \q 4$\times$10$^{2}$-8$\times$10$^{3}$
\lsun; while still low, the upper end of this range
approaches the bolometric luminosity derived from the SED fitting.

Considering the uncertainties in luminosity estimates and in the EGO 24 \um\/ flux
densities, we adopt the following
approach: For every EGO in our VLA sample with a 24 \um\/ flux density
in \citetalias{maserpap}, we calculate  
\begin{math} L_{bol}(24 \mu m) = S_{24 \mu m} * 10^{-26} * \Delta \nu * \chi * 4\pi D^{2} * 2.48 \times 10^{12}
\end{math} where L$_{bol}$(24 $\mu$m) has units of \lsun, S$_{24 \mu m}$ is the 24 \um\/ flux density from
\citetalias{maserpap} in Jy, $\Delta \nu =$2.8521 $\times$10$^{12}$ Hz is the bandwidth of
the MIPS 24 \um\/ filter \citep{Cohen09}, $\chi$ is 8 or 150
(corresponding to the range in F$_{bol}$/F$_{MIPS 24 \mu m}$ discussed
above), and D is the distance to the source in kpc.  
Similarly, L$_{\nu}$(8 GHz) is defined as 
4$\pi$D$^{2}$S$_{3.6 cm}$.
These values are plotted in
Figure~\ref{hoare_plots} along with data 
from
\citet{HoareFranco07}.  In addition, data
for low-mass sources from Table 5 of \citet{Anglada98} are plotted
for comparison.  Each EGO is plotted twice, with the upper estimate
for L$_{bol}$(24 \um) plotted as a filled
triangle, and the lower estimate for L$_{bol}$(24 \um) plotted as an open triangle for the same value of L$_{\nu}$(8 GHz).
For EGOs with multiple cm counterparts coincident with MIPS 24 \um\/ emission (G35.03+0.35 and G49.27$-$0.34), all are plotted in
Figure~\ref{hoare_plots} (sets of triangles at the same L$_{bol}$(24 \um) but
different L$_{\nu}$(8 GHz)).  For nondetections, L$_{\nu}$(8 GHz) is calculated for
the 4$\sigma$ 3.6 cm upper limit.  We consider G11.92$-$0.61
a 3.6 cm nondetection and plot the 4$\sigma$ upper limit, rather
than extrapolating from the 1.3 cm detection assuming a spectral
index.
All discussions of luminosities of course depend on distance estimates.  
For our EGO sample, we adopt the kinematic distances from
\citetalias{maserpap}, which are based on the new Galactic rotation
model of \citet{Reid09}.

Several salient points emerge from an
examination of Figure~\ref{hoare_plots}.  (1) None of the EGOs extend
into the upper right corner of the plot populated by UC HII regions.  In particular, G35.03+0.35-CM1,
for which we have good constraints on the electron density that place
it firmly within the UC HII category (\S\ref{g3503}), has L$_{\nu}$(8 GHz)\q
10$^{13}$ \whz , towards the lower end of the HC HII range.  G35.03+0.35-CM1 is
less of an outlier in the line width plots of \citet[Fig. 6]{Hoare07}.  In line
width/size parameter space, the small recombination line width
\citep[17.6 \kms,][]{rsro} and size (\q 0.007 pc) of
G35.03+0.35-CM1 are consistent with the region occupied by UC
HIIs, though G35.03+0.35-CM1 has a narrower line width than any source
in the \citet{Hoare07} sample.  (2) EGOs detected at 3.6 cm generally lie in, or slightly
above, the region of L$_{\nu}$(8 GHz) versus L$_{bol}$(24 \um) parameter space occupied by
winds and ionized jets.  The EGO detections to some extent
fill in the gap between HC HIIs and wind/jet sources in the
original \citet{HoareFranco07} plot.  
However, as noted above, G35.03+0.35-CM1 and
G49.27$-$0.34-CM1 fall in the HC HII
region of Figure~\ref{hoare_plots}, and their classification as HCs is questionable. 
(3) The EGO nondetection
upper limits all fall roughly in the region 
occupied by wind/jet sources, several orders of magnitude in
L$_{\nu}$(8 GHz) below the \citet{HoareFranco07} UC and HC HII regions and more than an order of magnitude below the UC HII G35.03+0.35-CM1.
(4) Even with highly uncertain, and possibly low, estimates for the
bolometric luminosities, the properties of the EGO population as a whole
are consistent with MYSO wind/jet sources and HC HIIs, 
and distinct from the low-mass
sources from the \citet{Anglada98} sample.

In sum, our data definitively show that the vast majority
(12/14\q86\%) of EGOs in our VLA sample are \emph{not} associated with
UC HII regions.  The two EGOs that are associated with UC/C HII
regions (G35.03+0.35 and G49.27$-$0.34) also have other cm-$\lambda$
continuum counterparts, which may be at earlier evolutionary stages.
The other EGO cm counterparts have spectral index limits consistent with either HC
HII regions or winds/jets; their weak 3.6 cm
emission, relative to bolometric luminosity, is similar to well-studied wind and jet sources.
In three
EGOs (G24.94+0.07, G28.83$-$0.25, and G35.03+0.35), faint 3.6 cm
emission is detected coincident with 6.7 GHz Class II \meth\/ masers,
indicating ionized gas very near the central MYSO.
One important exception to the general trend of intermediate
spectral-index emission 
is G11.92$-$0.61-CM1, which may be an
optically thick 
HC HII region.

\subsection{The Evolutionary State of EGOs}\label{discussion_evol}

As noted above, the majority of EGOs in our VLA survey
(8/14\q57\%) are not associated with detectable cm-$\lambda$ continuum
emission in our deep observations.  In evaluating the
significance of cm-$\lambda$ nondetections towards MYSOs, there is
always a degeneracy between age and mass: is the (proto)star not
emitting ionizing radiation because it is young, or because it is not,
and will never be, massive/luminous enough to do so?  This degeneracy
is to an important extent physical, and not simply an observational
limitation: as a MYSO grows by accretion (whether competitively or
otherwise), it will go through a stage at which its mass and
luminosity are moderate, rapid accretion is ongoing, and its final
mass is indeterminant, though limited by the mass reservoir of its natal clump.

There is increasing evidence, from both modeling and observations, that accretion from a large-scale reservoir is a crucial element of the massive star formation process, and a key difference between clustered massive and isolated low-mass star formation \citep[e.g.][and references therein]{Roberto09,Peters11mag,Wang10,Roberto10}.
So one way of observationally addressing the final mass of an actively accreting (proto)star, albeit indirectly, is to look at the mass of the clump-scale reservoir available for ongoing accretion.  All but one of the EGOs in our VLA sample fall within the area of the 1.1 mm Bolocam Galactic Plane Survey \citep[BGPS;][]{Aguirre11}, and are associated with 1.1 mm sources in the BGPS catalog \citep{Rosolowsky10}.\footnote{For the one source that falls outside the BGPS survey area, G11.92$-$0.61, we calculate gas masses as described in the text but using the 850 \um\/ flux density of 12 Jy from \citet{Walsh03}, R=100, and $\kappa_{\nu}$= 2.2 (see also \citetalias{C11}).}  We calculate gas masses from the 1.1 mm dust continuum emission   
\begin{equation}
M_{gas}= {\frac{4.79 \times 10^{-14} R S_{\nu}(Jy) D^2(kpc)}{B(\nu,T_{dust})\kappa_{\nu}}},
\label{dust_mass_eqn}
\end{equation}
where S$_{\nu}$ is the
flux density from the BGPS catalog corrected by the recommended factor
of 1.5$\pm$0.15 \citep{Aguirre11,Dunham10}, D is the distance to the
source, B($\nu,T_{dust}$) is the Planck function, R is the gas-to-dust
mass ratio, and $\kappa_{\nu}$ is the dust mass opacity coefficient in
units of cm$^{2}$ g$^{-1}$.  We follow recent BGPS studies
\citep[e.g.][]{Dunham10,Dunham11} in adopting $\kappa_{271
GHz}$/R=0.0114 cm$^{2}$ g$^{-1}$ (assuming a gas-to-dust mass ratio of
100).  
For most of our sample, no information about the clump-scale dust or
gas temperature is available.  We adopt two limiting dust
temperatures, 16 and 28 K, bracketing the range (22$\pm$6 K) from
\citet{Dunham11} for ``Group 3'' star-forming BGPS sources (those
identified with either an EGO or young RMS source).
Clump masses calculated using equation~\ref{dust_mass_eqn} and the integrated flux density of the
associated BGPS source are listed in Tables~\ref{nondetect_table} and \ref{detect_table}.
For three clumps
that correspond to ATLASGAL 870 \um\/ sources, the ATLASGAL mass \citep[derived using T$_{k}$ estimates
from \ammonia\/ observations;][]{atlasgal} falls within our range of estimated
clump gas masses.

We do not attempt to correct the clump mass estimates for free-free contributions to the 1.1 mm flux density for
several reasons: (1) Many BGPS sources are large, and
either encompass area outside the VLA 1.3 cm primary beam and/or
extended cm sources for which flux
densities cannot be measured from our data.  Hence we cannot correct for free-free
contributions in a uniform way for all sources in our sample. (2) The
uncertainties in the clump mass estimates are dominated by the uncertainty
in the correction factor for BGPS catalog fluxes ($\pm$15\%) and 
in the assumed dust temperature.  For the sources in our sample with
the strongest free-free emission, G28.28$-$0.36 and G49.27$-$0.34, the
total contribution of all cm sources within the VLA 1.3 cm primary
beam corresponds to changes in the mass estimates of \q7\% and $<$1\%, respectively, assuming the steepest spectral indices allowed by our
data ($\Delta$M/M \q130/1840 \msun\/ and \q 45/5050 \msun\/ for T$_{dust}$=16 K, within the uncertainties quoted in
Table~\ref{detect_table}).  

Figure~\ref{mass_histo}a shows histograms of masses estimated using
T$_{dust}$=16 K and the integrated flux density from the BGPS catalog
for EGOs with/without cm-$\lambda$ counterparts.  For comparison,
Figure~\ref{mass_histo}b shows masses estimated using the 80\pp\/
aperture fluxes from the BGPS catalog.  This provides a comparison of
the mass at a uniform angular scale across our EGO sample, though the
apertures are centered on the 1.1 mm peaks, and may be offset from the
EGO positions.  No trend in clump masses for EGOs with/without
cm-$\lambda$ counterparts is seen in Figure~\ref{mass_histo}a or b.
Adopting T$_{dust}$=28 K for all sources does not change the relative
positions of cm detections/nondetections, only shifts the absolute
mass scale.  If we instead assume that sources with cm-$\lambda$
continuum detections are warmer, the mass range of clumps associated
with EGO cm detections and nondetections still largely overlaps
(Fig.~\ref{mass_histo}c,d).

For BGPS clumps associated with HII regions, \citet{Katharine09} found
a correlation between the clump mass and that of the ionizing stars.
This indicates a relationship between the mass of a clump, and the
mass of the massive stars that it forms.  This idea, combined with the
importance of accretion from large-scale reservoirs to massive star
formation, suggests that the EGOs in our survey sample have the
potential to continue accreting, and eventually become massive stars
that will ionize their surroundings (i.e.\ form UC HII regions).  As
shown in Figure~\ref{mass_histo}, there is no apparent trend in clump
masses for EGOs with cm-$\lambda$ detections as compared to
cm-$\lambda$ nondetections, as might be expected if the
presence/absence of cm continuum emission was related to
reservoir--and so to final stellar--mass.  In this picture, the
paucity of cm-$\lambda$ continuum detections towards EGOs is a measure
of their youth.


Additional support for this explanation is derived
from the SED fitting of G19.01$-$0.03, a cm nondetection.  This MYSO is actively driving a bipolar outflow, traced by Class I \meth\/ masers, SiO emission, and high-velocity, collimated \co\/ and \hco\/ lobes \citepalias{maserpap,C11}.
An active outflow implies active accretion, which is expected
to alter the properties of the central (proto)star, affecting its output of ionizing radiation.
Based on the SED modeling, the central
(proto)star of G19.01$-$0.03 has a luminosity of \q 10$^{4}$ \lsun,
mass of $\gtrsim$ 10 \msun, and is accreting at a rate of \q 10$^{-3}$
\msun\/ yr$^{-1}$ \citepalias{C11}.  Figure~\ref{g19_rad_temp_plot}a
shows the stellar radius and temperature for the best-fit models
($\chi^{2}-\chi^{2}_{best}$ per datapoint $<$3) to the SED composed of
the IRAC, MIPS, and SMA data from \citetalias{C11}.  Both swollen, very low temperature (\q
5000 K) and hotter ($>$10$^{4}$ K) models are allowed; however, even
the highest stellar temperature in Figure~\ref{g19_rad_temp_plot}a
falls below the 30,000 K threshold for ``significant'' Lyman continuum
output suggested by \citet{HoareFranco07}.
Figure~\ref{g19_rad_temp_plot}b shows that if the low angular
resolution (single dish) (sub)mm flux density estimates at 870 \um\/
and 1.1 mm \citep{atlasgal,Rosolowsky10} are substituted for the SMA
data, all the well-fit models are pushed to large radii and low
temperatures.\footnote{For both plots, SEDs were fit assuming a minimum uncertainty for all flux densities of 10\% to mirror recent single-dish studies \citep[e.g.][]{Pandian10}, which differs somewhat from the errors assumed by \citetalias{C11}.}.

\citet{Longmore09} report the detection of spatially extended,
optically thin free-free emission in the fields of six 6.7
GHz \meth\/ masers previously reported as 3.6 cm continuum nondetections \citep{Walsh98}, and caution that the
evolutionary state of \meth\/ maser sources may be misconstrued if
extended free-free emission is missed in high-resolution
interferometric observations.  We believe that our cm continuum study of
EGOs is robust against such misclassification for several reasons.
(1) The sensitivity and imaging fidelity of our VLA observations are
better than the ATCA survey of \citet{Walsh98}; though we cannot
recover fluxes for sources more extended than \q 10\pp, our data are
sensitive to structures up to \q 20\pp\/ in angular scale
(\S\ref{vlaobs}).
This is illustrated by our 3.6 cm detection of the HII region F
G28.83-0.25-CM1, which was missed by \citet{Walsh98}. 
(2) We have examined
GLIMPSE and MIPSGAL 24 \um\/ images for the \citet{Longmore09} targets, and the extended HII regions
reported manifest themselves as structured,
multiband IRAC sources and/or 8 \um\/ nebulae associated with bright
24 \um\/ emission \citep[see also discussion of
``diffuse red clumps'' by][]{Battersby10}.  This is consistent with
the MIR morphologies of our ``field'' cm sources
(Fig.~\ref{3color_big}), including those overresolved in our
observations (and so candidates for being possibly ``extended and
overlooked'').  (3)  In several cases, the extended
HII regions reported by \citet{Longmore09} are offset from the \meth\/
maser positions, and the masers are more closely
associated with HC HII regions or with 3 mm emission attributed to
dust.  So, as \citet{Longmore09} note, the presence of more evolved
HII regions in the vicinity of 6.7 GHz \meth\/ masers may simply be due to
the clustered nature of massive star formation.  In fact, we see
several potential examples of this in our data. 
In these
cases, the high-resolution GLIMPSE and MIPSGAL MIR images are helpful
in identifying evolved sources and assessing their likely
relation to (less evolved) sources of interest
(\S\ref{individ_sources}).

\section{Conclusions}\label{conclusions}

Our exceptionally deep VLA 3.6 cm survey (median rms \q 30 \ujb)
provides convincing evidence that most EGOs are young MYSOs, prior to
the development of a UC HII region. No continuum emission is detected
at 3.6 or 1.3 cm (median rms \q 0.25 \mjb) towards the majority
(8/14 \q 57\%) of EGOs in our survey sample.  Only two sources (2/14 \q 14\%)
are associated with optically thin cm-$\lambda$ emission from UC/C HII regions
(one of which is the only one of our targets without an associated 6.7
GHz \meth\/ maser). Each of these EGOs exhibits cm-$\lambda$ multiplicity, with
morphological evidence that a less-evolved cm continuum source may be
the driving source of the outflow traced by 4.5 \um\/ and Class I \meth\/
maser emission. With the exception of the UC HII regions, most of the
cm-$\lambda$ EGO counterparts are weak ($\lesssim$ 1 mJy), unresolved at the \q 1\pp\/
resolution of our survey, undetected at 1.3 cm, and characterized by
intermediate spectral indices consistent with either HC HII regions or
ionized winds or jets. Three of these weak, unresolved 3.6 cm sources
are coincident with 6.7 GHz \meth\/ maser emission, indicative of
ionized gas very near the central (proto)star.  Likely to be an
optically thick HC HII region, one EGO cm counterpart is detected only
at 1.3 cm and is associated with a hot core and \water\/ and 6.7 GHz \meth\/
maser emission.  The predominance of very weak cm-$\lambda$ continuum
detections in our survey indicates the promise of the greatly enhanced
continuum sensitivity of the Expanded Very Large Array (EVLA) for
understanding the early stages of massive star formation, including
the earliest appearance of ionized gas around accreting MYSOs.

In sum, the cm wavelength properties of EGOs are indicative of an
early stage of massive star formation, when photoionizing feedback
from the central MYSO is not yet a significant influence on the
(proto)cluster environment.  The presence of 4.5 \um\/ emission, Class I
\meth\/ masers, and SiO emission \citepalias{maserpap,C11}
indicates that these EGOs are actively driving outflows, and so
(presumably) actively accreting.  Notably, two of the EGOs with weak
or no cm emission in our survey (G11.92$-$0.61 and G19.01$-$0.03) have
collimated, high-velocity molecular outflows, in contrast to some
recent theoretical predictions that ionizing radiation will dominate
outflow dynamics in massive star-forming regions \citep[e.g.][]{Peters11mag}. 
Already sufficiently massive to excite 6.7 GHz \meth\/ maser
emission, the surveyed EGOs are associated with significant clump-scale gas reservoirs.
These clumps can provide sufficient material for
sustained, rapid accretion as the EGOs continue to grow from MYSOs into
massive stars.

\acknowledgments

This research has made use of NASA's Astrophysics Data System
Bibliographic Services and the SIMBAD database operated at CDS,
Strasbourg, France.  Support for this work was provided by NSF grant AST-0808119.
C.J.C. was partially supported during this work by a National Science
Foundation Graduate Research Fellowship, and is currently supported by
an NSF Astronomy and Astrophysics Postdoctoral Fellowship under award
AST-1003134.  C.J.C. thanks C. Purcell for providing the CORNISH aperture photometry program and for helpful discussions, J. Mottram for providing the calibration factor for MIPS 24 \um\/ flux to bolometric luminosity calculated from the RMS sample, M. Hoare for providing the data files for his published plots, and T. Robitaille for helpful discussions about SED fitting.

{\it Facilities:}  \facility{VLA ()}, \facility{Spitzer ()}


\begin{deluxetable}{cccccccc}
\tablewidth{0pt}
\tablecaption{Observational Parameters\label{obstable}}
\tablehead{ 
\colhead{Source Name} & 
\multicolumn{2}{c}{J2000.0 Coordinates\tablenotemark{a}} & 
\colhead{Phase Cal.} &
\colhead{Syn. Beam\tablenotemark{b}} & 
\colhead{3.6 cm} &
\colhead{1.3 cm} & 
\colhead{uv} \\
\colhead{} &
\colhead{$\alpha$ ($^{\rm h}~~^{\rm m}~~^{\rm s}$)} &
\colhead{$\delta$ ($^{\circ}~~{\arcmin}$~~\pp)} &
\colhead{(1.3 cm)} &
\colhead{\pp $\times$ \pp [\degree]} &
\colhead{rms} & 
\colhead{rms} & 
\colhead{range?}\\
\colhead{} & 
\colhead{} & 
\colhead{} & 
\colhead{} & 
\colhead{} &
\colhead{(\mjb)} & 
\colhead{(\mjb)}  & 
\colhead{(k$\lambda$)}
}
\tablecolumns{8}
\tabletypesize{\scriptsize}
\setlength{\tabcolsep}{0.05in}
\startdata
G10.29$-$0.13 & 18 08 49.3 & $-$20 05 57.3 &  1820-254 &1.45$\times$0.86[0.38]
& 0.136 & 0.32 & $>$10 (3.6cm)  \\
G11.92$-$0.61 & 18 13 58.1 & $-$18 54 16.7 &  1820-254 & 1.45$\times$0.85[1.51]
& 0.052 & 0.25 & \nodata\\
G18.67+0.03 & 18 24 53.7 & $-$12 39 20.0 & 1832-105 &1.25$\times$0.91[-2.03]
& 0.031 & 0.23 & \nodata\\
G18.89$-$0.47 & 18 27 07.9 & $-$12 41 35.5 &1832-105  & 1.25$\times$0.91[-3.77]
&0.037  &0.25 & $>$25 (3.6cm)\\
G19.01$-$0.03 & 18 25 44.8 & $-$12 22 45.8 &1832-105  & 1.24$\times$0.93[-3.72]
& 0.029 &0.26 & $>$10 (3.6cm)\\
G19.36$-$0.03 & 18 26 25.8 & $-$12 03 56.9 & 1832-105 & 1.24$\times$0.92[-4.07]
& 0.029 & 0.26 & \nodata\\
G22.04+0.22 & 18 30 34.7 & $-$9 34 47.0 &1832-105  & 1.20$\times$0.95[-0.24]
 &0.031  & 0.27 & \nodata \\
G24.94+0.07 & 18 36 31.5 & $-$07 04 16.0 & 1832-105  &1.14$\times$0.95[-1.39]
& 0.036 & 0.25 & $>$10 (1.3cm)  \\
G25.27$-$0.43 & 18 38 56.9 & $-$07 00 48.0 & 1832-105 &1.16$\times$0.96[-1.88]
& 0.030 & 0.26 & \nodata \\
G28.28$-$0.36 & 18 44 13.2 & $-$04 18 04.0 & 1851+005  &1.12$\times$0.95[6.75]
& 0.053 & 0.24 & \nodata \\
G28.83$-$0.25 & 18 44 51.0 & $-$03 45 49.0 & 1851+005  &1.09$\times$0.95[7.52]
& 0.029 & 0.23& $>$10 (3.6,1.3cm)  \\
G35.03+0.35 & 18 54 01.0 & +02 01 20.0 & 1851+005 & 1.05$\times$0.96[8.92]
& 0.028 & 0.20& \nodata \\
G49.27$-$0.34 & 19 23 06.7 & +14 20 13.0 & 1924+156 &0.96$\times$0.89[-7.35]
& 0.135 & 0.18 & \nodata \\
G49.42+0.33 & 19 20 59.3 & +14 46 48.0 & 1924+156& 0.93$\times$0.88[-17.83]
& 0.027 & 0.17& \nodata \\
\enddata
\tablenotetext{a}{Phase tracking centers of the VLA observations; in some cases, these may differ slightly from the \citetalias{egocat} positions.}
\tablenotetext{b}{At 1.3 cm; the 3.6 cm data were convolved to this resolution, see \S\ref{vlaobs}.}
\end{deluxetable}

\begin{deluxetable}{rllcccllccc}
\rotate
\tablewidth{0pt}
\tablecaption{EGO Detections: Observed Properties of Centimeter Continuum Sources\label{egos_table_detect}}
\tablehead{ 
\colhead{Source Name} & 
\multicolumn{5}{c}{\underline{3.6 cm}} &
\multicolumn{5}{c}{\underline{1.3 cm}}\\
\colhead{} & 
\multicolumn{2}{c}{J2000.0 Coordinates\tablenotemark{a}} & 
\colhead{Peak Intensity\tablenotemark{b}} & 
\colhead{S$_{\nu}$\tablenotemark{c}} & 
\colhead{Size\tablenotemark{c}} &
\multicolumn{2}{c}{J2000.0 Coordinates\tablenotemark{a}} & 
\colhead{Peak Intensity\tablenotemark{b}} & 
\colhead{S$_{\nu}$\tablenotemark{c}} & 
\colhead{Size\tablenotemark{c}}\\
\colhead{}& 
\colhead{$\alpha$ ($^{\rm h}~~^{\rm m}~~^{\rm s}$)} &
\colhead{$\delta$ ($^{\circ}~~{\arcmin}$~~\pp)} &
\colhead{(\mjb)} & 
\colhead{(mJy)} & 
\colhead{(\pp $\times$ \pp [P.A. \degree])} &
\colhead{$\alpha$ ($^{\rm h}~~^{\rm m}~~^{\rm s}$)} &
\colhead{$\delta$ ($^{\circ}~~{\arcmin}$~~\pp)} &
\colhead{(\mjb)} & 
\colhead{(mJy)} & 
\colhead{(\pp $\times$ \pp [P.A. $^{\circ}$])} 
}
\tablecolumns{11}
\tabletypesize{\scriptsize}
\setlength{\tabcolsep}{0.05in}
\startdata
EGO G11.92$-$0.61-CM1 & \nodata & \nodata & $<$0.21 & \nodata & \nodata & 18 13 58.10 & $-$18 54 20.5 & 1.07(0.25) & \nodata & \nodata \\ 
EGO G24.94+0.07-CM1 & 18 36 31.572 & $-$07 04 16.93 & 0.53(0.04) & \nodata & \nodata & \nodata & \nodata & $<$1.01 & \nodata & \nodata \\ 
EGO G28.28$-$0.36-CM1 &  18 44 13.33 & $-$04 18 04.3 & 0.21(0.05) &  \nodata & \nodata & \nodata & \nodata & $<$0.96 & \nodata & \nodata \\ 
EGO G28.83$-$0.25-CM1 & 18 44 50.737 & $-$03 45 49.26 & 0.33(0.03)  &  \nodata & \nodata & \nodata & \nodata & $<$0.92 & \nodata & \nodata \\
-CM2 & 18 44 51.08 & $-$03 45 48.3 & 0.14(0.03) & \nodata & \nodata & \nodata & \nodata & $<$0.92 & \nodata & \nodata \\
EGO G35.03+0.35-CM1 & 18 54 00.49098 & +02 01 18.292 & 12.34(0.03) & 13.76(0.05) & 0.36$\times$0.32[111] & 18 54 00.4902 & +02 01 18.327 & 10.30(0.20) & 12.07(0.38) & 0.51$\times$0.31[157]\\
-CM2 & 18 54 00.6498 & +02 01 19.32 & 1.29(0.03) & 1.49(0.05) & 0.51$\times$0.24[21] & 18 54 00.647 & +02 01 19.53 & 2.88(0.19) & \nodata & \nodata \\
-CM3 & 18 54 00.766 & +02 01 22.82 & 0.45(0.03) & \nodata & \nodata & \nodata & \nodata & $<$0.81 & \nodata & \nodata \\  
-CM4 & 18 54 00.523 & +02 01 15.60 & 0.23(0.03) &  \nodata & \nodata & \nodata & \nodata & $<$0.81 & \nodata & \nodata \\  
-CM5 & 18 54 00.45 & +02 01 21.2 & 0.13(0.03) &  \nodata & \nodata & \nodata & \nodata & $<$0.81 & \nodata & \nodata \\ 
EGO G49.27$-$0.34-CM1\tablenotemark{d} &  19 23 06.91 & +14 20 16.7 & 2.44(0.14) & 73(9) & 9.5 & 19 23 06.91  & +14 20 16.7  & 2.63(0.18) &  67(7) & 9.1 \\ 
-CM2 & 19 23 06.61 &  +14 20 12.0 & 0.61(0.14) &  \nodata & \nodata & \nodata & \nodata & $<$0.71 & \nodata & \nodata \\  
\enddata
\tablenotetext{a}{The number of significant figures indicates the
formal uncertainty in the centroid position obtained from a
two-dimensional Gaussian fit, or (for sources $<$5 $\sigma$), a
one-pixel uncertainty in the position of the peak (see also
\S\ref{results}).}  
\tablenotetext{b}{All upper limits are 4$\sigma$.
For detections, the quoted error is the formal error from the two
dimensional Gaussian fit, or the 1$\sigma$ rms in the image, depending
on whether the source was detected with sufficient signal-to-noise to
fit (see \S\ref{results}).} 
\tablenotetext{c}{Measured from
two-dimensional Gaussian fits; formal errors from the fits are given
in parentheses.}  
\tablenotetext{d}{Source properties measured using polygonal aperture
photometry program developed for the CORNISH survey.  The quoted
uncertainty in the peak intensity is the 1$\sigma$ rms in the image.
The quoted size is the equivalent source diameter, defined as the
diameter of a disk with the same solid angle as the source polygon.
See also \S\ref{results}.}
\end{deluxetable}

\begin{deluxetable}{rcc}
\tablewidth{0pt}
\tablecaption{EGO Nondetections: Observed Centimeter Limits \label{egos_table_nondetect}}
\tablehead{ 
\colhead{Source Name} & 
\colhead{\underline{3.6 cm}} &
\colhead{\underline{1.3 cm}}\\
\colhead{} & 
\colhead{Peak Intensity\tablenotemark{a}} & 
\colhead{Peak Intensity\tablenotemark{a}} \\
\colhead{} & 
\colhead{(\mjb)} & 
\colhead{(\mjb)} 
}
\tablecolumns{3}
\tabletypesize{\scriptsize}
\setlength{\tabcolsep}{0.05in}
\startdata
EGO G10.29$-$0.13 & $<$0.54\tablenotemark{b} &  $<$1.27   \\
EGO G18.67+0.03 &  $<$0.12 &  $<$0.94  \\
EGO G18.89$-$0.47 &  $<$0.15 &  $<$0.98   \\
EGO G19.01$-$0.03 &  $<$0.12 &  $<$1.04  \\
EGO G19.36$-$0.03 &  $<$0.12 &  $<$1.03 \\
EGO G22.04+0.22 & $<$0.12 &  $<$1.06  \\
EGO G25.27$-$0.43 & $<$0.12 &  $<$ 1.03  \\
EGO G49.42+0.33 &  $<$0.11 & $<$0.68 \\
\enddata
\tablenotetext{a}{All upper limits are 4$\sigma$.
} 
\tablenotetext{b}{The noise level is
elevated due to image artifacts caused by copious
complex, extended emission in the field (and beyond the half-power
point of the primary beam) that is poorly imaged in our
observations.}
\end{deluxetable}

\begin{deluxetable}{rllcccllcccc}
\rotate
\tablewidth{0pt}
\tablecaption{Observed Properties of Additional Centimeter Continuum Sources in EGO Fields\label{others_table}}
\tablehead{ 
\colhead{Source Name} & 
\multicolumn{5}{c}{\underline{3.6 cm}} &
\multicolumn{5}{c}{\underline{1.3 cm}} &
\colhead{Ang. Sep.\tablenotemark{d}}\\
\colhead{} & 
\multicolumn{2}{c}{J2000.0 Coordinates\tablenotemark{a}} & 
\colhead{Peak Intensity\tablenotemark{b}} & 
\colhead{S$_{\nu}$\tablenotemark{b}} & 
\colhead{Size} &
\multicolumn{2}{c}{J2000.0 Coordinates\tablenotemark{a}} & 
\colhead{Peak Intensity\tablenotemark{b,c}} & 
\colhead{S$_{\nu}$} & 
\colhead{Size}& 
\colhead{from EGO}\\
\colhead{}& 
\colhead{$\alpha$ ($^{\rm h}~~^{\rm m}~~^{\rm s}$)} &
\colhead{$\delta$ ($^{\circ}~~{\arcmin}$~~\pp)} &
\colhead{(\mjb)} & 
\colhead{(mJy)} & 
\colhead{} &
\colhead{$\alpha$ ($^{\rm h}~~^{\rm m}~~^{\rm s}$)} &
\colhead{$\delta$ ($^{\circ}~~{\arcmin}$~~\pp)} &
\colhead{(\mjb)} & 
\colhead{(mJy)} & 
\colhead{} & 
\colhead{(\pp)}
}
\tablecolumns{11}
\tabletypesize{\scriptsize}
\setlength{\tabcolsep}{0.05in}
\startdata
F G10.29$-$0.13-CM1 & 18 08 50.6 & $-$20 06 00 & \tablenotemark{e} &\tablenotemark{e} & $\gtrsim$21\pp &  18 08 50.6 & $-$20 06 00 & \tablenotemark{e} & \tablenotemark{e} & $\gtrsim$21\pp & $\gtrsim$10\\
F G18.67+0.03-CM1 & 18 24 52.6056 & $-$12 39 20.005 & 4.29 (0.03) & 6.35(0.07) & 0.83 $\times$ 0.63 [41]\tablenotemark{f} & 18 24 52.607 & $-$12 39 20.02 & 3.7(0.4) & \nodata & \nodata & 16\\ 
F G19.36$-$0.03-CM1 & 18 26 25.628 & $-$12 03 49.43 & 0.42(0.03) & 0.99(0.09) & 1.39$\times$1.06[55]\tablenotemark{f} & \nodata & \nodata & $<$1.03  & \nodata & \nodata & 8\\
-CM2 & 18 26 26.379 & $-$12 04 19.90 & 0.48(0.03) & \nodata & \nodata &  \nodata & \nodata & $<$1.16 & \nodata & \nodata & 24\\ 
F G25.27$-$0.43-CM1 & 18 38 59.6635 & $-$07 00 55.15 & 1.02(0.03) & \nodata & \nodata &   \nodata & \nodata & $<$ 1.23  & \nodata & \nodata & 40 \\ 
-CM2 & 18 38 57.576 & $-$07 01 14.78 & 0.31(0.03) &   \nodata & \nodata &   \nodata & \nodata & $<$ 1.16  & \nodata & \nodata & 28\\ 
F G28.28$-$0.36-CM1\tablenotemark{g} & 18 44 15.11 & $-$04 17 56.1 & 77.32(0.05)  & 576(3) & 7.7\pp & 18 44 15.11  & -04 17 56.2 & 69.6(0.3) &  510(11) & 8.9\pp & 30 \\
-CM2 & 18 44 09.802 & $-$04 18 00.12 & 1.51(0.07) & 2.0(0.1) & 0.69$\times$0.54[42]\tablenotemark{f} & \nodata & \nodata & $<$1.47 &  \nodata & \nodata & 51 \\
-CM3 & 18 44 16.498 & $-$04 18 08.86 & 1.03(0.06) & 3.0(0.2) & 1.53$\times$1.33[176]\tablenotemark{f} & \nodata & \nodata & $<$1.69 &  \nodata & \nodata & 50 \\
F G28.83$-$0.25-CM1 & 18 44 51.3 & $-$03 45 22 &  \tablenotemark{e} &  \tablenotemark{e} & \q27\pp & \nodata & \nodata & \nodata & \nodata & \nodata & $\gtrsim$13\\
F G49.42+0.33-CM1 & 19 20 59.81 &  14 46 49.4 & 0.14(0.03) & \nodata & \nodata & \nodata & \nodata &  $<$0.68 & \nodata & \nodata & 11 \\
-CM2& 19 20 56.4 & 14 46 44 & \tablenotemark{e}  & \tablenotemark{e}  & \q 20\pp & \nodata & \nodata & \nodata & \nodata & \nodata & $\gtrsim$30 \\
\enddata
\tablenotetext{a}{Measured as described in \S\ref{results} for different source types.  For sources fit with Gaussians, the number of significant figures indicates the formal uncertainty in the centroid position from the Gaussian fit.}
\tablenotetext{b}{
For detections, the quoted error is the formal error from the two
dimensional Gaussian fit, or the 1$\sigma$ rms in the image (depending
on whether the source was detected with sufficient signal-to-noise to
fit, \S\ref{results}), except as otherwise noted.}
\tablenotetext{c}{All upper limits are 4$\sigma$, where $\sigma$ is the rms at the location of the source in the primary-beam-corrected image (for sources far from the pointing center, this will differ from the rms at the EGO position).} 
\tablenotetext{d}{Angular separation between the centimeter emission and the EGO position from \citetalias{egocat} (minimum angular separation for extended cm sources).}
\tablenotetext{e}{Extended source poorly imaged by the interferometer, so no fluxes are reported.}
\tablenotetext{f}{Deconvolved source size (\pp $\times$ \pp [P.A. $^{\circ}$]) as determined by fitting a single 2D Gaussian component.}
\tablenotetext{g}{Source properties measured using polygonal aperture
photometry program developed for the CORNISH survey.  The quoted
uncertainty in the peak intensity is the 1$\sigma$ rms in the image.
The quoted size is the equivalent source diameter, defined as the
diameter of a disk with the same solid angle as the source polygon.
See also \S\ref{results}.}

\end{deluxetable}

\begin{deluxetable}{ccccccccccccc}
\tablewidth{0pt}
\tablecaption{EGO Cm Nondetections: Derived Physical Properties and Limits \label{nondetect_table}}
\tablehead{ 
\colhead{Source Name} & 
\colhead{Distance\tablenotemark{a}} &
\multicolumn{3}{c}{\underline{3.6 cm}\tablenotemark{b}} &
\multicolumn{3}{c}{\underline{1.3 cm}\tablenotemark{b}} &
\multicolumn{2}{c}{Clump mass\tablenotemark{c}} &
\colhead{L$_{\nu}$(8 GHz)} &
\multicolumn{2}{c}{L$_{bol}$(24 \um )\tablenotemark{d}}\\
\colhead{} & 
\colhead{(kpc)} & 
\colhead{l\tablenotemark{e}} & 
\colhead{l\tablenotemark{e}} & 
\colhead{N$_{Lyc}$\tablenotemark{f}} & 
\colhead{l\tablenotemark{e}} & 
\colhead{l\tablenotemark{e}} & 
\colhead{N$_{Lyc}$\tablenotemark{f}} & 
\colhead{T$_{d}$=16K} & 
\colhead{T$_{d}$=28K} & 
\colhead{(W Hz$^{-1}$)} & 
\colhead{min} &
\colhead{max} 
\\
\colhead{} & 
\colhead{} & 
\colhead{(mpc)} & 
\colhead{(AU)} & 
\colhead{(s$^{-1}$)} & 
\colhead{(mpc)} & 
\colhead{(AU)} & 
\colhead{(s$^{-1}$)} & 
\colhead{(\msun)} & 
\colhead{(\msun)} & 
\colhead{} & 
\colhead{(\lsun)} &
\colhead{(\lsun)}
\\
\colhead{} & 
\colhead{} & 
\colhead{} & 
\colhead{} & 
\colhead{$\times$10$^{44}$} & 
\colhead{} & 
\colhead{} & 
\colhead{$\times$10$^{45}$} & 
\colhead{} & 
\colhead{} & 
\colhead{} & 
\colhead{}\\ 
}
\tablecolumns{12}
\tabletypesize{\scriptsize}
\setlength{\tabcolsep}{0.05in}
\startdata
EGO G10.29$-$0.13 & 2.19 & $<$0.39 & $<$80 & $<$2.5 & $<$0.22 & $<$46 & $<$0.7 & 1220$^{+220}_{-200}$ & 570$^{+100}_{-100}$ & $<$3.1$\times$10$^{11}$ & \nodata & \nodata\\
\\
EGO G18.67+0.03 & 4.98 & $<$0.42 & $<$86 & $<$2.8 & $<$0.44 & $<$90 & $<$2.5 & 1660$^{+340}_{-310}$ & 780$^{+160}_{-140}$ & $<$3.6$\times$10$^{11}$ & 6.9$\times$10$^{2}$ & 1.3$\times$10$^{4}$\\
\\
EGO G18.89$-$0.47 & 4.49 & $<$0.42 & $<$86 & $<$2.9 & $<$0.40 & $<$83 & $<$2.1 & 5350$^{+950}_{-870}$ & 2520$^{+450}_{-410}$ & $<$3.6$\times$10$^{11}$ &  7.2$\times$10$^{1}$ & 1.3$\times$10$^{3}$\\
\\
EGO G19.01$-$0.03 & 4.20 & $<$0.35 & $<$72 & $<$2.0 & $<$0.39 & $<$80 & $<$2.0 & 1140$^{+220}_{-200}$ & 540$^{+110}_{-100}$ & $<$2.5$\times$10$^{11}$ &  4.4$\times$10$^{2}$ & 8.2$\times$10$^{3}$\\
\\
EGO G19.36$-$0.03 & 2.43 & $<$0.20 & $<$42 & $<$0.7 & $<$0.22 & $<$46 & $<$0.7 & 1090$^{+190}_{-180}$ & 510$^{+90}_{-80}$ & $<$8.5$\times$10$^{10}$ & \nodata & \nodata \\
\\
EGO G22.04+0.22 & 3.62 & $<$0.30 & $<$62 & $<$1.5 & $<$0.34 & $<$70 & $<$1.5 & 1650$^{+300}_{-280}$ & 780$^{+140}_{-130}$ & $<$1.9$\times$10$^{11}$ & 3.1$\times$10$^{2}$ & 5.8$\times$10$^{3}$\\
\\
EGO G25.27$-$0.43 & 3.86 & $<$0.32 & $<$66 & $<$1.7 & $<$0.36 & $<$73 & $<$1.6 & 960$^{+200}_{-180}$ & 450$^{+100}_{-90}$ & $<$2.1$\times$10$^{11}$ & 2.1$\times$10$^{1}$ & 4.0$\times$10$^{2}$\\
\\
EGO G49.42+0.33 & 12.29 & $<$0.98 & $<$202 & $<$16 & $<$0.92 & $<$189 & $<$11 & 3440$^{+900}_{-800}$ & 1620$^{+430}_{-380}$ & $<$2.0$\times$10$^{12}$ & \nodata & \nodata \\
\enddata
\tablenotetext{a}{From \citetalias{maserpap}.}
\tablenotetext{b}{All quantities are upper limits corresponding to a 4$\sigma$ nondetection in our VLA images.}
\tablenotetext{c}{Clump masses calculated from the integrated flux density measurement in the BGPS catalog \citep{Rosolowsky10}, multipled by a factor of 1.5$\pm$0.15 as recommended by \citet{Aguirre11}.  The quoted uncertainties include the uncertainty in this multiplicative factor.  See \S\ref{discussion_evol}.}
\tablenotetext{d}{Estimated from the 24 \um\/ flux density from \citetalias{maserpap}, see \S\ref{lir_dis}.  \nodata indicates no 24 \um\/ flux density in \citetalias{maserpap}, either because a clear counterpart is not present, or because confusion/blending precludes measuring a reliable flux for the EGO counterpart.}
\tablenotetext{e}{Twice the radius calculated in equation~\ref{thick_radius_limit} (\S\ref{phy_prop_results}).  Assumes optically thick emission.}
\tablenotetext{f}{Assumes optically thin emission, \S\ref{phy_prop_results}.}
\end{deluxetable}

\begin{deluxetable}{rccccccccccccc}
\rotate
\tablewidth{0pt}
\tablecaption{EGO Cm Detections: Derived Physical Properties and Limits \label{detect_table}}
\tablehead{ 
\colhead{Source Name} & 
\colhead{Distance\tablenotemark{a}} &
\colhead{$\alpha$\tablenotemark{b}} &
\multicolumn{3}{c}{\underline{3.6 cm}\tablenotemark{c}} &
\multicolumn{3}{c}{\underline{1.3 cm}\tablenotemark{c}} &
\multicolumn{2}{c}{Clump mass\tablenotemark{d}} &
\colhead{L$_{\nu}$(8 GHz)} &
\multicolumn{2}{c}{L$_{bol}$(24 \um )\tablenotemark{e}}\\
\colhead{} & 
\colhead{(kpc)} & 
\colhead{} & 
\colhead{l\tablenotemark{f}} & 
\colhead{l\tablenotemark{f}} & 
\colhead{N$_{Lyc}$\tablenotemark{g}} & 
\colhead{l\tablenotemark{f}} & 
\colhead{l\tablenotemark{f}} & 
\colhead{N$_{Lyc}$\tablenotemark{g}} & 
\colhead{T$_{d}$=16K} & 
\colhead{T$_{d}$=28K} & 
\colhead{(W Hz$^{-1}$)} & 
\colhead{min} &
\colhead{max} 
\\
\colhead{} & 
\colhead{} & 
\colhead{} & 
\colhead{(mpc)} & 
\colhead{(AU)} & 
\colhead{(s$^{-1}$)} & 
\colhead{(mpc)} & 
\colhead{(AU)} & 
\colhead{(s$^{-1}$)} & 
\colhead{(\msun)} & 
\colhead{(\msun)} & 
\colhead{} & 
\colhead{(\lsun)} &
\colhead{(\lsun)}
\\
\colhead{} & 
\colhead{} & 
\colhead{} & 
\colhead{} & 
\colhead{} & 
\colhead{$\times$10$^{44}$} & 
\colhead{} & 
\colhead{} & 
\colhead{$\times$10$^{45}$} & 
\colhead{} & 
\colhead{} & 
\colhead{} & 
\colhead{}\\ 
}
\tablecolumns{13}
\tabletypesize{\scriptsize}
\setlength{\tabcolsep}{0.05in}
\startdata
EGO G11.92$-$0.61-CM1 & 3.80 & $>$1.7 & $<$0.42 & $<$86 & $<$2.89 & 0.36 & 73 & 1.66 & 1100\tablenotemark{h} & 490\tablenotemark{h}  & $<$3.6$\times$10$^{11}$ & 1.1$\times$10$^{3}$ &  2.0$\times$10$^{4}$ \\
\\ 
EGO G24.94+0.07-CM1 & 2.99 & $<$0.7 & 0.52 & 108 & 4.51 & $<$0.27 & $<$56 & $<$0.97 & 340$^{+80}_{-70}$ & 160$^{+40}_{-40}$ & 5.7$\times$10$^{11}$ & 8.9$\times$10$^{1}$ &  1.7$\times$10$^{3}$ \\
\\
EGO G28.28$-$0.36-CM1 & 3.29 & $<$1.6 & 0.36 & 75 & 2.16 & $<$0.29 & $<$60 & $<$1.12 & 1840$^{+320}_{-300}$ & 870$^{+150}_{-140}$ & 2.7$\times$10$^{11}$ & \nodata & \nodata \\
\\
EGO G28.83$-$0.25-CM1 & 5.03 & $<$1.1 & 0.70 & 143 & 7.95 & $<$0.44 & $<$90 & $<$2.51 &  3340$^{+600}_{-550}$ & 1580$^{+280}_{-260}$ & 1.0$\times$10$^{12}$ & \nodata & \nodata \\
\\
-CM2 &                  5.03 & $<$1.9 & 0.45 & 93 & 3.37 & $<$0.44 & $<$90 & $<$2.51 &  3340$^{+600}_{-550}$ & 1580$^{+280}_{-260}$ & 4.2$\times$10$^{11}$ & 6.5$\times$10$^{2}$ &  1.2$\times$10$^{4}$ \\
\\
EGO G35.03+0.35-CM1 & 3.43 & $-$0.1 & \nodata \tablenotemark{i} & \nodata \tablenotemark{i} & 154 & \nodata \tablenotemark{i} & \nodata \tablenotemark{i} & 15.3 & 2070$^{+360}_{-330}$ & 980$^{+170}_{-160}$ &  1.9$\times$10$^{13}$ & 1.5$\times$10$^{3}$ &  2.8$\times$10$^{4}$ \\
\\
-CM2 &                 3.43 & 0.7 & 1.01 & 208 & 16.7 & 0.53 & 109 & 3.65 & 2070$^{+360}_{-330}$ & 980$^{+170}_{-160}$& 2.1$\times$10$^{12}$ & 1.5$\times$10$^{3}$ &  2.8$\times$10$^{4}$ \\
\\
-CM3 &                  3.43 &  $<$0.6 & 0.55 & 114 & 5.04 & $<$0.28 & $<$58 & $<$1.03 &  2070$^{+360}_{-330}$ & 980$^{+170}_{-160}$& 6.3$\times$10$^{11}$ & 1.5$\times$10$^{3}$ &  2.8$\times$10$^{4}$ \\
\\
-CM4 &                  3.43 &  $<$1.3 & 0.40 & 82 & 2.58 & $<$ 0.28 & $<$ 58 & $<$ 1.03 &  2070$^{+360}_{-330}$ & 980$^{+170}_{-160}$& 3.2$\times$10$^{11}$ &   1.5$\times$10$^{3}$ &  2.8$\times$10$^{4}$ \\
\\    
-CM5 &                  3.43 & $<$1.9 & 0.30 & 61 & 1.46 & $<$0.28 & $<$58 & $<$1.03 & 2070$^{+360}_{-330}$ & 980$^{+170}_{-160}$& 1.8$\times$10$^{11}$ & 1.5$\times$10$^{3}$ &  2.8$\times$10$^{4}$ \\
\\
EGO G49.27$-$0.34-CM1 & 5.55 & $-$0.1 & \nodata \tablenotemark{i} & \nodata \tablenotemark{i} & 2140 & \nodata \tablenotemark{i} & \nodata \tablenotemark{i} & 222 & 5050$^{+900}_{-830}$ & 2380$^{+430}_{-390}$ &2.7$\times$10$^{14}$ &  2.4$\times$10$^{3}$ &  4.5$\times$10$^{4}$ \\
\\
-CM2 &                  5.55 & $<$0.2 & \nodata \tablenotemark{i} & \nodata \tablenotemark{i} & 17.9 & \nodata \tablenotemark{i} & \nodata \tablenotemark{i} & $<$2.35 &  5050$^{+900}_{-830}$ & 2380$^{+430}_{-390}$& 2.2$\times$10$^{12} $ & 2.4$\times$10$^{3}$ &  4.5$\times$10$^{4}$ \\

\enddata
\tablenotetext{a}{From \citetalias{maserpap}.}
\tablenotetext{b}{Spectral index $\alpha$, S$_{\nu} \propto \nu^{\alpha}$, calculated between 3.6 and 1.3 cm from our VLA data.}
\tablenotetext{c}{All limits correspond to 4$\sigma$ upper limits in our VLA images.}
\tablenotetext{d}{Clump masses calculated from the integrated flux density measurement in the BGPS catalog \citep{Rosolowsky10}, multipled by a factor of 1.5$\pm$0.15 as recommended by \citet{Aguirre11}.  The quoted uncertainties include the uncertainty in this multiplicative factor.  See \S\ref{discussion_evol}.}
\tablenotetext{e}{Estimated from the 24 \um\/ flux density from \citetalias{maserpap}, see \S\ref{lir_dis}.  \nodata indicates no 24 \um\/ flux density in \citetalias{maserpap}, either because a clear counterpart is not present, or because confusion/blending precludes measuring a reliable flux for the EGO counterpart.}
\tablenotetext{f}{Twice the radius calculated in equation~\ref{thick_radius_limit} (\S\ref{phy_prop_results}).  Assumes optically thick emission.}
\tablenotetext{g}{Assumes optically thin emission, \S\ref{phy_prop_results}.}
\tablenotetext{h}{G11.92$-$0.61 is the only EGO in our sample that falls outside the BGPS coverage.  The masses listed here were calculated from the 850 \um\/ SCUBA flux reported by \citet{Walsh03} (12 Jy), assuming T$_{dust}$ as for the BGPS sources and $\kappa_{850 \mu m}$=2.2 \citep[interpolated from the values tabulated by][]{OH94}.}
\tablenotetext{i}{Not calculated, since the radius estimate assumes unresolved optically thick emission (\S\ref{phy_prop_results}), and the spectral index indicates that the emission is optically thin.} 
\end{deluxetable}

\newpage

\begin{figure}
\plotone{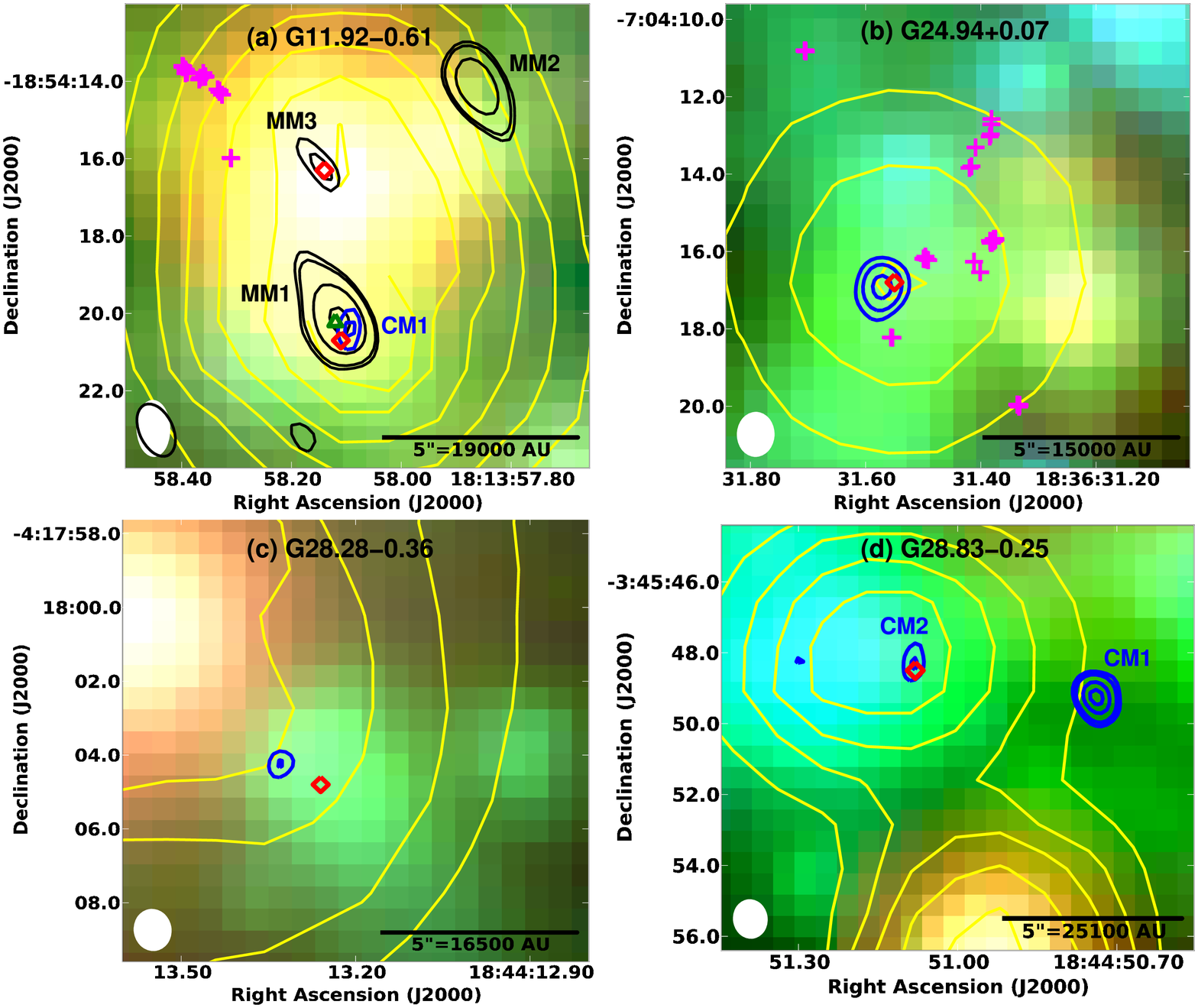}
\caption{}
\label{3color_zoom}
\end{figure}

\begin{figure}
\addtocounter{figure}{-1}
\plotone{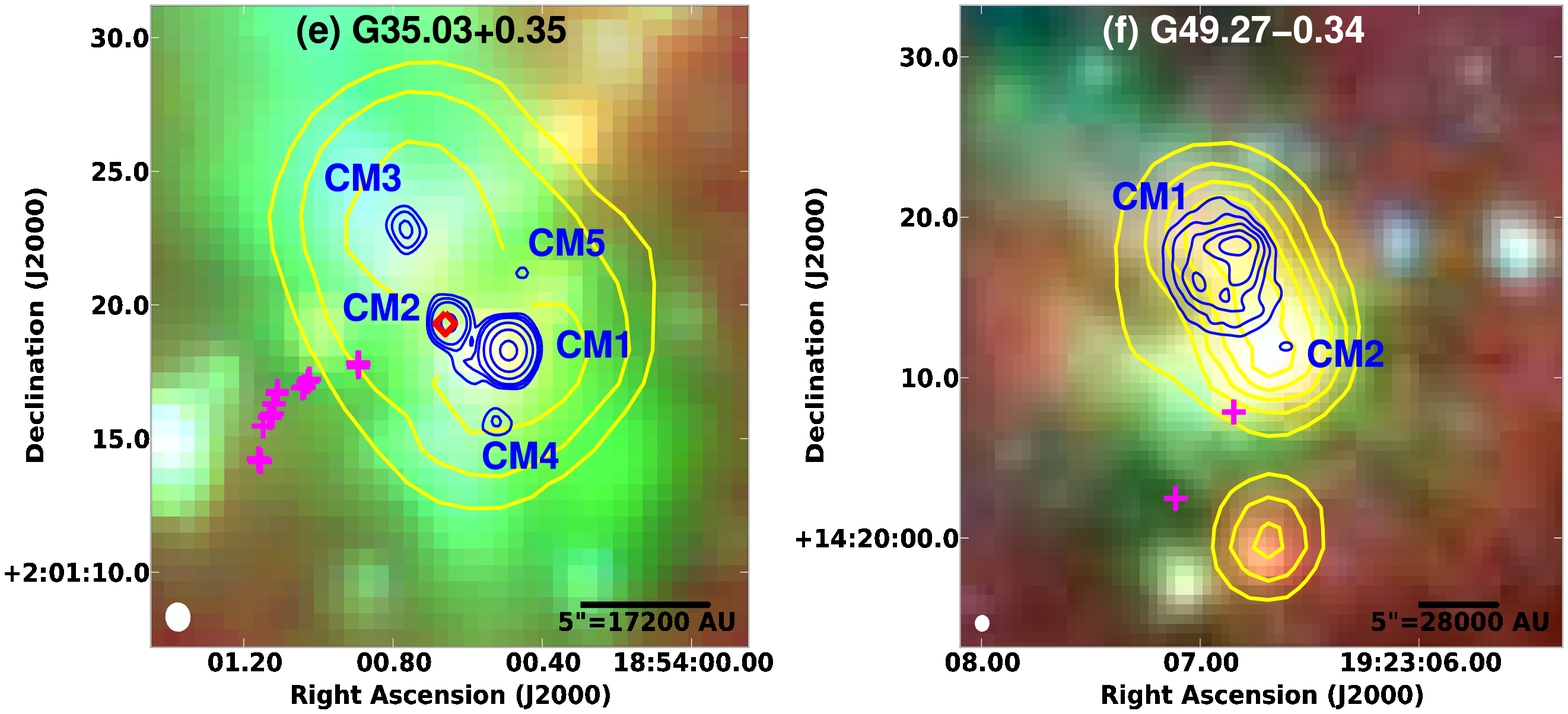}
\caption{}
\end{figure}

\begin{figure}
\addtocounter{figure}{-1}
\caption{\textbf{Cm-$\lambda$ EGO detections.}
Three-color \emph{Spitzer} GLIMPSE images (RGB: 8.0, 4.5, 3.6 \um)
overlaid with contours of VLA cm continuum emission (blue) and MIPS 24
\um\/ emission (yellow).  All panels are centered on the VLA pointing
center except (e), which is centered on the position of
G35.03+0.35-CM2.  Positions of 6.7 GHz Class II and 44 GHz Class I
\meth\/ masers from \citetalias{maserpap} are plotted as red diamonds and
magenta crosses, respectively.  The VLA beam is shown at lower left in
each panel (Table~\ref{obstable}).  The VLA images shown have not
been corrected for the primary beam, but all measurements were made
from corrected images.  (a) G11.92$-$0.61: VLA 1.3 cm continuum
contours at (3,4)$\times \sigma=$ 2.54E-4 \jb\/ and CARMA 1.4 mm
continuum (black contours), levels (4,5,10,20) $\times \sigma =$ 4.3
\mjb\/ (the CARMA beam is shown at lower left).  The \water\/ maser
from \citet{HC96} is plotted as a green triangle.  MIPS 24 \um\/
contour levels: 600,900,1200,1500,1800,2100 MJy sr$^{-1}$ (b)
G24.94+0.07: VLA 3.6 cm continuum contour levels (4,7,12)$\times
\sigma=$ 3.57E-5 \jb .  MIPS 24 \um\/ contour levels: 300,600,900 MJy
sr$^{-1}$ (c) G28.28$-$0.36: VLA 3.6 cm continuum contour levels
(3,4)$\times \sigma=$5.25E-5 \jb .  MIPS 24 \um\/ contour levels:
900,1200,1500,1800 MJy sr$^{-1}$ (d) G28.83$-$0.25: VLA 3.6 cm
continuum contour levels (3.5,4.5,7,9)$\times \sigma=$ 2.92E-5 \jb
. MIPS 24 \um\/ contour levels: 600,900,1200,1500,1800 (e)
G35.03+0.35: VLA 3.6 cm contour levels (4,7,12,36,108,324)$\times
\sigma=$ 2.78E-5 \jb . MIPS 24 \um\/ contour levels: 900,1200,1800 MJy
sr$^{-1}$ (f) G49.27$-$0.34: VLA 3.6 cm contour levels (4,7,10,13,16)
$\times \sigma=$ 1.35E-4 \jb .  MIPS 24 \um\/ contour
levels:600,900,1200,1500,1800 MJy sr$^{-1}$.  
}
\end{figure}

\begin{figure}
\plotone{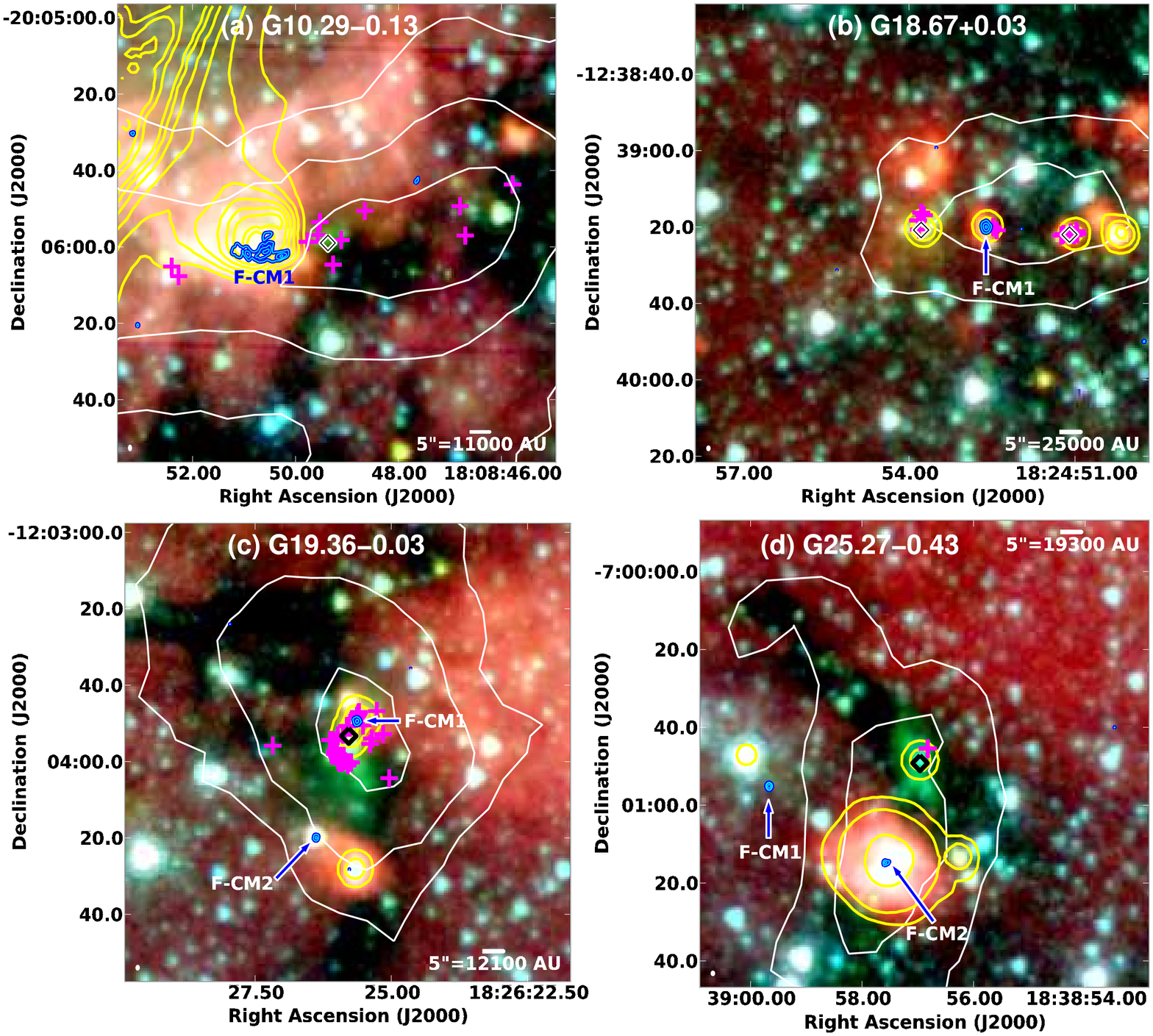}
\caption{}
\label{3color_big}
\end{figure}

\begin{figure}
\plotone{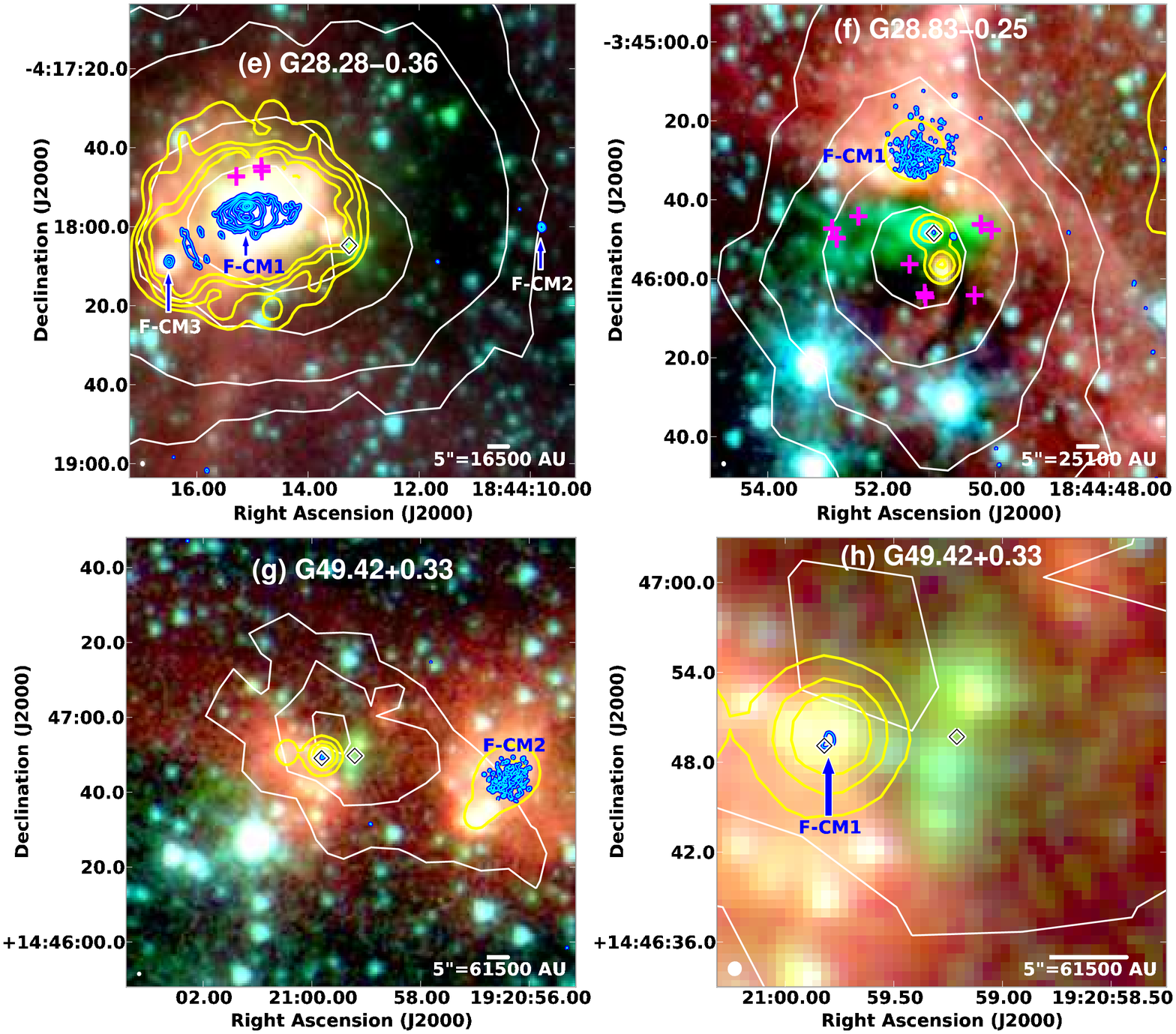}
\addtocounter{figure}{-1}
\caption{}
\end{figure}

\begin{figure}
\addtocounter{figure}{-1}
\caption{\textbf{Field Cm-$\lambda$ Continuum Sources.}
Three-color \emph{Spitzer} GLIMPSE images (RGB: 8.0, 4.5, 3.6 \um)
overlaid with contours of VLA 3.6 cm continuum emission (blue/cyan),
BGPS 1.1 mm emission (white), and MIPS 24 \um\/ emission (yellow).
All panels are centered on the VLA pointing center.  
Positions of 6.7 GHz Class II and 44 GHz Class I \meth\/ masers from
\citetalias{maserpap} are plotted as diamonds and magenta crosses, respectively.  The VLA beam is shown 
at lower left in each panel
(Table~\ref{obstable}).  
The VLA
images shown here have not been corrected for the primary beam, but all
measurements were made from corrected images.  
(a) G10.29$-$0.13: VLA 3.6 cm continuum contour levels:
(4,7,10)$\times \sigma =$ 1.36E-4 \jb ; BGPS 1.1 mm contour levels: (10,40,80) $\times
\sigma =$ 0.02 \jb ; MIPS 24 \um\/ contour levels: 900,1200,1500,1800,2100,2400 MJy
sr$^{-1}$ 
(b) G18.67+0.03: VLA 3.6 cm continuum contour levels: (4,40,120) $\times \sigma =$
3.06E-5 \jb ; BGPS 1.1 mm contour levels: (10,20) $\times \sigma =$ 0.03 \jb ; MIPS 24 \um\/ contour levels: 600,1200,1800 MJy sr$^{-1}$ 
(c) G19.36$-$0.03: VLA 3.6 cm continuum contour levels: (4,12)$\times
\sigma =$2.93E-5 \jb ; BGPS 1.1 mm contour levels: (10,20,40) $\times \sigma =$ 0.03 \jb ;
MIPS 24 \um\/ contour levels: 600,1300,2000 MJy sr$^{-1}$ 
(d) G25.27$-$0.43: VLA 3.6 cm continuum contour levels:
(4,7,21)$\times \sigma =$ 2.95E-5\jb ; BGPS 1.1 mm contour levels: (10,20) $\times \sigma
=$ 0.018 \jb ; MIPS 24 \um\/ contour levels: 75,150,600 MJy sr$^{-1}$ 
(e) G28.28$-$0.36:
VLA 3.6 cm continuum contour levels: (4,10,20,50,150,250,450,650,850,1250)$\times \sigma = $
5.25E-5 \jb ; BGPS 1.1 mm contour levels: (10,20,50,80)$\times \sigma = $ 0.015 \jb ; MIPS 24
\um\/ contour levels : 900,1200,1500,1800 MJy sr$^{-1}$ 
(f) G28.83$-$0.25: VLA 3.6 cm continuum contour levels:
(4,5,7)$\times \sigma = $2.86E-5 \jb ;
BGPS 1.1 mm contour levels: (10,20,50,100)$\times \sigma = $ 0.015 \jb ; MIPS 24 \um\/ contour levels:
600,1200,1800 MJy sr$^{-1}$ 
(g) G49.42+0.33: VLA 3.6 cm continuum contour levels: (4,5,7)$\times
\sigma = $ 2.69E-5 \jb ; BGPS 1.1 mm contour levels: (5,10,15)$\times \sigma = $ 0.025 \jb
; MIPS 24 \um\/ contour levels: 150,300,450 MJy sr$^{-1}$ 
(h) Same as (g); only the 4 and 5 $\sigma$ VLA 3.6 cm contours are visible within
the smaller field of view.
The BGPS rms is a function of Galactic longitude \citep[][Fig. 11]{Aguirre11}.
}
\end{figure}

\begin{figure}
\plottwo{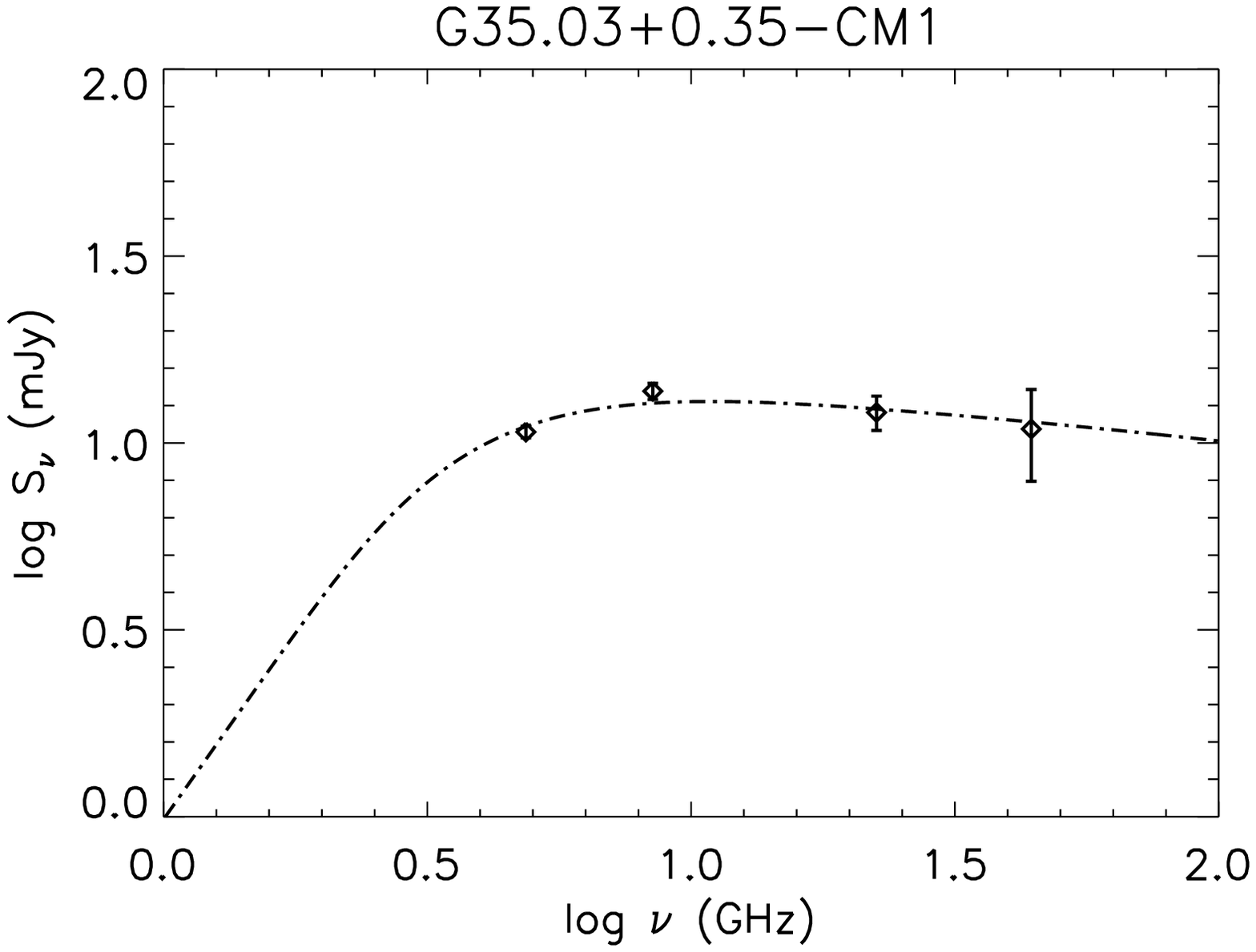}{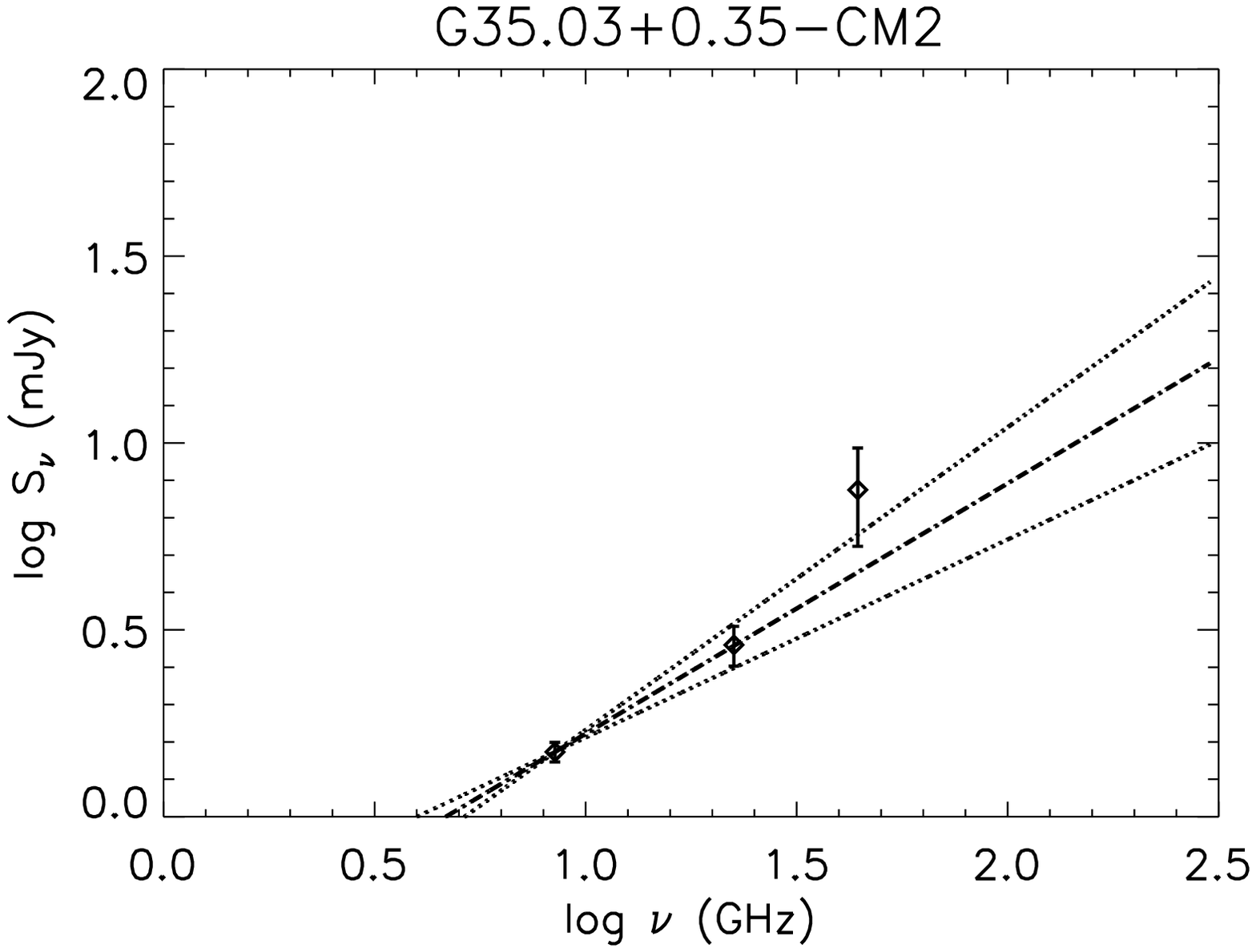}
\caption{SEDs for the cm sources EGO G35.03+0.35-CM1 (left) and G35.03+0.35-CM2 (right).  The dot-dashed line in the left panel is the fit described in \S\ref{g3503}.  In the right panel, lines of $\alpha=$0.67 (dot-dashed) and of $\alpha \pm$ 1$\sigma=$0.14 (dotted), extrapolated from the 3.6 cm flux density, are shown (\S\ref{g3503}.)}
\label{g3503_seds}
\end{figure}

\begin{figure}
\epsscale{0.9}
\plotone{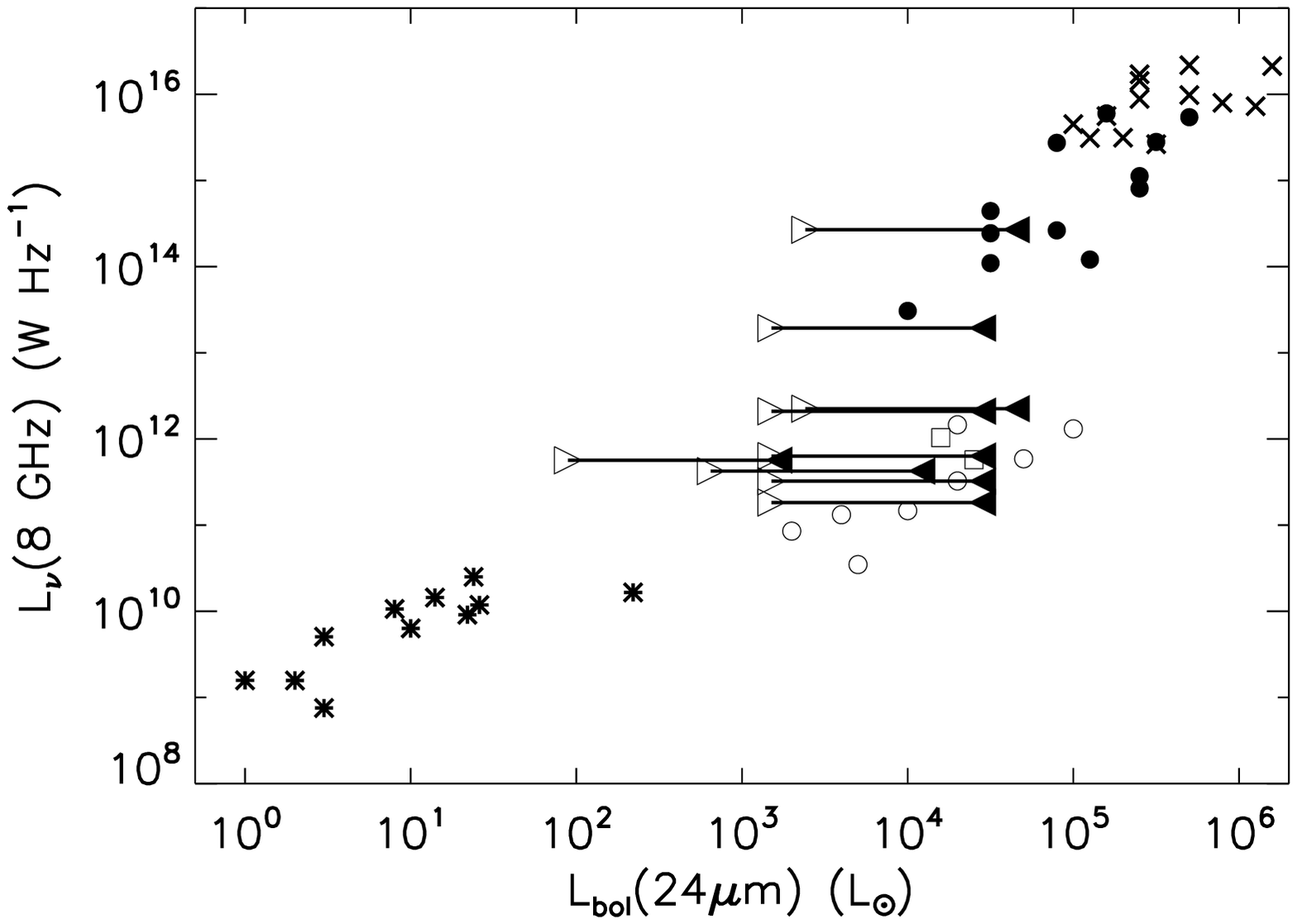}\\
\plotone{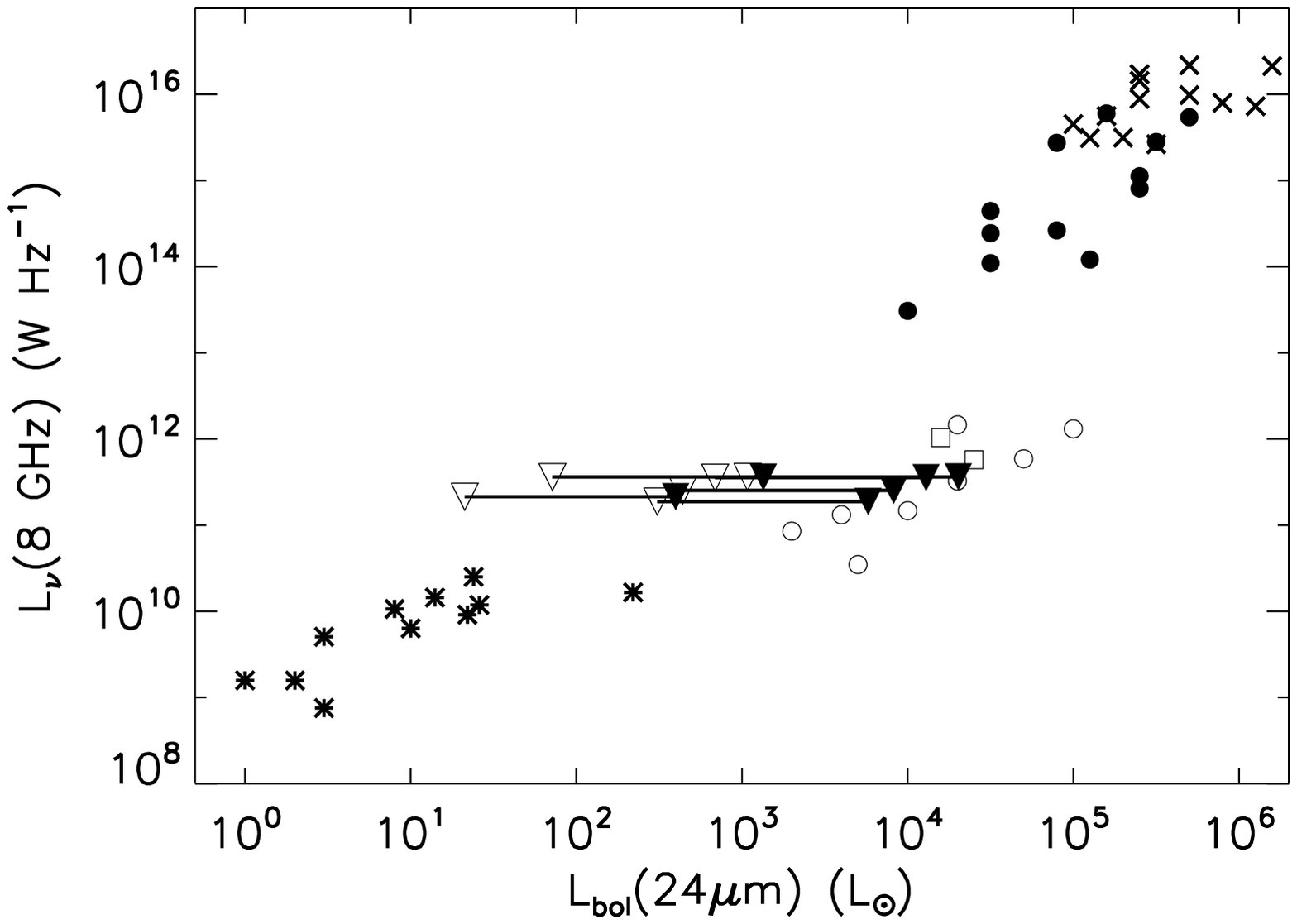}
\caption{}
\label{hoare_plots}
\end{figure}

\begin{figure}
\addtocounter{figure}{-1}
\caption{Plots of L$_{\nu}$(8 GHz) v. bolometric luminosity L$_{bol}$(24 \um) for EGOs
detected at 3.6 cm (top) and for EGO nondetections (bottom).  UC HIIs,
HC HIIs, wind, and jet sources from \citet{HoareFranco07} are plotted
as X's, solid dots, open circles, and open squares, respectively;
low-mass sources from \citet{Anglada98} are plotted as stars.  For
each EGO cm counterpart detected, the datapoint corresponding to our
lower estimate for L$_{bol}$(24 \um) is plotted as a right-facing open triangle,
and the datapoint corresponding to our upper estimate for L$_{bol}$(24 \um) is
plotted as a left-facing solid triangle, with the two connected with a line (see \S\ref{lir_dis}).  
Nondetections are
plotted as downward-pointing triangles, with a solid triangle for the upper and an open triangle for the lower L$_{bol}$(24 \um)
estimate, joined by a line.  
}
\end{figure}

\begin{figure}
\epsscale{1.0}
\plotone{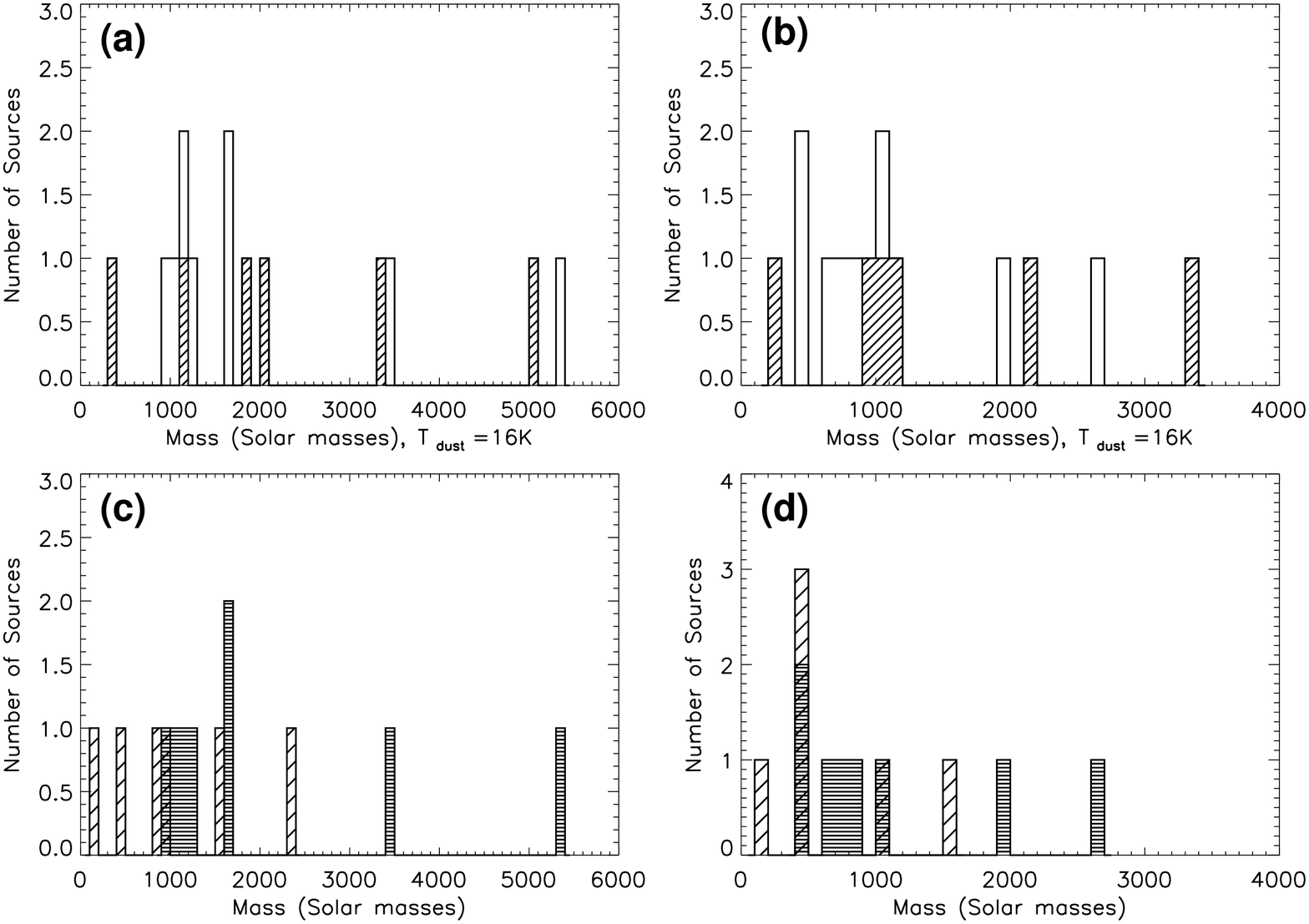}
\caption{\textbf{Top:} Histograms of the masses of clumps associated with EGOs calculated from the integrated flux densities (a) and 80\pp\/ aperture flux densities (b) in the BGPS catalog \citep{Rosolowsky10} for a dust temperature of 16 K.  For clarity, only nominal mass values are plotted.  Histograms of all clump masses (EGO cm detections and nondetections) are plotted as solid lines, clumps associated with EGO cm detections are plotted as hatched histograms.  \textbf{Bottom:}  Same as top, except T$_{dust}$=28 K is used for EGOs with cm detections.  Cm detections and nondetections are plotted as diagonally and horizontally hatched histograms, respectively. 
}
\label{mass_histo}
\end{figure}


\begin{figure}
\plotone{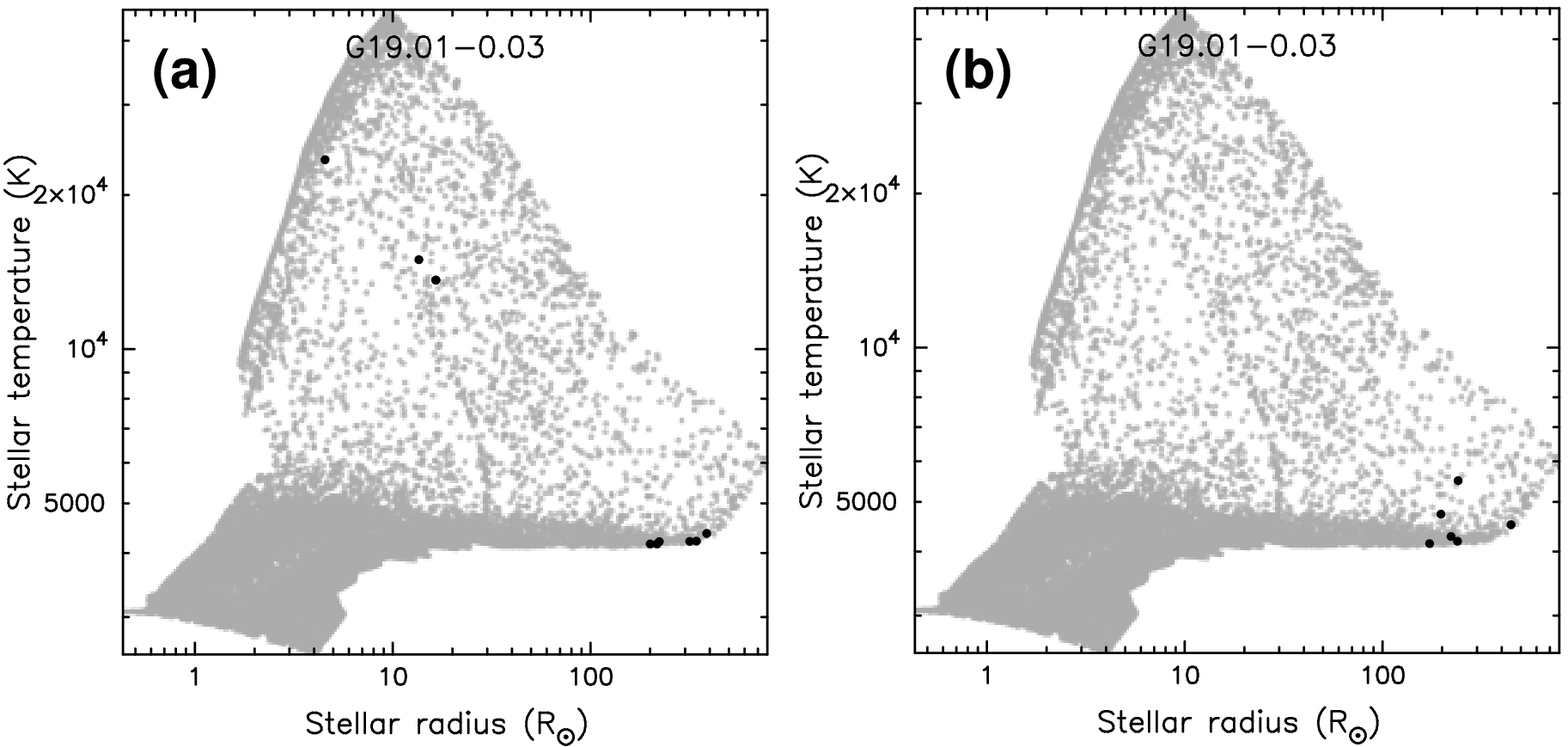}
\caption{Stellar radius and temperature for models well-fit ($\chi^{2}-\chi^{2}_{best}$ per datapoint $<$3) to the G19.01$-$0.03 SED using the \citet{Robitaille06,Robitaille07} model grid and fitter (black dots; the greyscale shows all models in the grid).  (a) Fits to the datapoints used in \citetalias{C11} (IRAC, MIPS 24 and 70 \um, and SMA 1.3 mm); (b) fits to the same MIR datapoints but with ATLASGAL 870 \um\/ and BGPS 1.1 mm data in place of the SMA 1.3 mm point.   }
\label{g19_rad_temp_plot}
\end{figure}

\end{document}